%% file: main.tex
\documentclass{iopjournal}

\usepackage{amsmath}
\usepackage{natbib}

\begin{document}

\articletype{Paper}

\title{Evaluating the prediction of subgrid eddy kinetic energy from a machine learning model}

\author{Thomas Wilder$^1$, Ranjini Swaminathan$^2$, David Storkey$^3$, and Till Kuhlbrodt$^1$}

\affil{$^1$Department of Meteorology and National Centre for Atmospheric Science, University of Reading, Reading, UK}

\affil{$^2$Department of Meteorology and National Centre for Earth Observation, University of Reading, Reading, UK}

\affil{$^3$Met Office, FitzRoy Road, Exeter, UK}

\email{thomas.wilder@ncas.ac.uk}


\begin{abstract}
Global climate model resolution is still too coarse to fully resolve mesoscale processes, therefore subgrid closures are required to approximate the feedback of eddies onto the resolved flow. With high-resolution datasets now readily available, there has been a rapid growth in the use of machine learning to develop new eddy closures that could help reduce model biases and uncertainty. In this work we employ an offline learning approach with a convolutional neural network to predict subgrid eddy kinetic energy ($KE_e$), laying the initial groundwork for a machine-learned scale-aware GEOMETRIC eddy closure for the UK's Earth System Model. We address the question of whether global parameterizations are needed or if regional closures are sufficient by evaluating  a number of neural network models through region and feature ablation experiments. ML parameterization models generalize to unseen regions when their training data share similar feature and target distributions with those regions. Models that fail to generalize incur distributional shift, or employ different filtering scales for the true eddy kinetic energy field. When holding out features, there is improvement among the generalizable ML models, with some feature spaces leading to improvements in some regions but degrading performance in others. Generally, power spectral density of $KE_e$ tend to diverge at small scales, though some feature spaces are able to minimize this divergence. Overall, the best performing model trained on all regions achieves time-mean and anomaly R2 scores of $R2\approx 0.95$ and $R2\approx 0.8-0.9$, respectively.
\end{abstract}

\input{introduction.tex}

\input{methods-and-theory.tex}

\input{results.tex}

\input{conclusion.tex}

\clearpage

\ack{The work of TW, RS, and TK was supported by the AI4PEX project (the UK Research and Innovation (UKRI) under the UK government’s Horizon Europe funding guarantee under grant agreement number 10114295, 10103109, and 10093450). TW and TK acknowledge the support of the European Union's Horizon 2020 research and innovation programme under grant agreement no. 101003536 (ESM2025 Earth System Models for the Future). RS received funding from the project Progressing Earth System Modelling for 
Tipping Point Early Warning Systems, PROMOTE (Grant number SCOP-PRO1-P027), part of 
the Forecasting Tipping Points programme of the Advanced Research + Invention Agency 
(ARIA). The high-resolution training data was produced on the ARCHER2 UK National Supercomputing Service (https://www.archer2.ac.uk). Model training and analysis was carried out on JASMIN (http://jasmin.ac.uk/), the UK collaborative data analysis facility. During the preparation of this manuscript, the authors used Claude Sonnet and Opus models to assist in writing scripts, producing figures and running sensitivity experiments. The authors reviewed, validated and critically assessed all AI-assisted outputs and take full responsibility for the content of the manuscript.}

\roles{TW: conceptualization, data curation, formal analysis, investigation, methodology, resources, software, validation, visualization, writing - original draft, review and editing. RS: conceptualization, investigation, methodology, project administration, supervision, writing - original draft, review and editing. DS: investigation, supervision, writing - review and editing. TK: conceptualization, funding acquisition, project administration.}

\data{The training data, the U-Net models and their predictions, along with the DINO $1/16^{\circ}$ namelist, are published on Zenodo \citep{wilderTrainingDataModels2026}. The figures, the data plotted in each figure, and the code that produces the data are published on Zenodo \citep{wilderFiguresFigureData2026}. The code used to build the training data, train the models and make the predictions is ml-parameterisation v1.0.0 found at https://github.com/thomaswilder/ml-parameterisation and archived on Zenodo \citep{wilderMlparameterisation2026}.}

\appendix

\input{appendix-A.tex}

\clearpage

\bibliographystyle{abbrvnat}
\bibliography{paper_1}

\end{document}

%% file: introduction.tex
\section{Introduction}

Successive generations of Earth System Models (ESMs) have improved in spatial resolution - going from tens of degrees to quarter degree in the latest generation of models from the Coupled Model Intercomparison Project (CMIP) \citep{dunneEvolvingCoupledModel2025}.
However, faithfully capturing mesoscale processes in current climate models is still a significant challenge, and one of utmost importance \citep{Hewitt2022}. In ocean processes, despite continued development of mesoscale parameterizations \citep{Bachmanetal2017, Mak2018, Jansen2019, Bachman2019}, many climate models still exhibit persistent biases in key climate metrics \citep{armourSeasurfaceTemperaturePattern2024, storkeyResolutionDependenceInterlinked2025}, contributing to large uncertainties in climate projections \citep{fox-kemperChallengesProspectsOcean2019}. For example, HadGEM3-GC31 MM, which uses the eddy-permitting NEMO ocean model ORCA025, exhibits warm and cold biases across the Southern Ocean and North Atlantic, negatively impacting the large scale circulation and contributing to improper regions of deep convection \citep{jacksonChallengesSimulatingAMOC2023}. Higher resolution models can counter some of these biases \citep{frigolaNorthAtlanticMean2025a}, but generally, computational limits still impose resolution constraints meaning explicit mesoscale processes must instead be parameterized. 

High-resolution data is now readily available \citep{clementCommunityGlobal12021} or simple to generate using idealized or realistic numerical simulations \citep{kammDINODiabaticModel2025a, seguraNextGEMSEnteringEra2025a}, which is suited for use in informing new subgrid closures. This data can then be leveraged by machine learning (ML) to drive new developments in subgrid eddy parameterizations, in particular for mesoscale eddies \citep{ZannaBolton2021}. The high fidelity data is typically pre-processed - filtered and coarsened - to a target horizontal resolution, such as $1/4$\textdegree, with the idea that the smaller scale information is now built into the coarser resolution. A supervised offline learning approach can be performed, where a mapping is learned from input features to a known target variable. In the case of machine learning subgrid closures for eddies, the target variable can be a potential vorticity or momentum forcing \citep{maulikSubgridModellingTwodimensional2019, boltonApplicationsDeepLearning2019, yanChoiceTrainingData2024, jakharLearningClosedFormEquations2024}, eddy kinetic energy or thickness fluxes in Gent-McWilliams (GM) \citep{GM1990} type schemes \citep{parteeUsingMachineLearning2022, balwadaDesignImplementationDataDriven2026}. Instead of developing a new scheme, recent approaches using reinforcement learning have been used to tune the unconstrained parameters within subgrid parameterizations \citep{nathReplacingTunableParameters2026}, which could be complementary within the ML closure space.

The application of machine learning to tackle challenges in climate change is leading to a new generation of Earth system models \citep{zannaFrameworkHybridPhysicsAI2025}. In particular, neural networks are able to utilize vast sums of data to learn complex physical relationships. Leveraging high-resolution numerical simulation data, \cite{maulikSubgridModellingTwodimensional2019} applied an artificial neural network (ANN) to learn a subgrid potential vorticity forcing term in a two-dimensional turbulence regime. The ANN was designed to take in a stencil of grid points that surround the grid point the target resides on. With this approach, their ML closure returned a $k^{-3}$ energy spectrum slope with greater energy at the grid scale compared with traditional LES subgrid schemes. \cite{boltonApplicationsDeepLearning2019} applied a convolutional neural network (CNN) to learn the subgrid momentum forcing term. The CNNs strength lies in its ability to pick out key features across the NNs hidden layers and relate them to the target. With this, they applied separate CNNs to learn a mapping from the non-divergent streamfunction to the $x$ and $y$ subgrid momentum forcing terms. The work of \cite{boltonApplicationsDeepLearning2019} has led to a number of follow up studies to the subgrid momentum forcing problem (e.g. \cite{perezhoginGenerativeDataDrivenApproaches2023, rossBenchmarkingMachineLearning2023}). Methods to improve the above mentioned sub-grid momentum forcing parameterization resulted in a more interpretable form using equation discovery methods \citep{zannaDataDrivenEquationDiscovery2020}. Since this work, several studies have improved this scheme further: filtering of the stress tensor to remove grid scale noise \citep{perezhoginStableImplementationDataDriven2024} and application of convolutional padding techniques to improve performance near coastlines \citep{zhangAddressingOutofSampleIssues2025}. Other lines of work have looked at improving the Gent-McWilliams type coefficient. \cite{balwadaDesignImplementationDataDriven2026} learned a thickness flux capable of generating positive eddy available potential energy tendency, while \cite{parteeUsingMachineLearning2022} embedded a surrogate model within a physics based parameterization, learning an eddy kinetic energy field from physically plausible features. Integrating machine learning within physics based schemes is considered a reasonable approach, ensuring some level of trust can remain with the black box neural network.

Physics based parameterizations are often derived using theory, observations, and numerical modeling, resulting in an equation based formula that is highly interpretable. Neural network based approaches on the other hand often tend to be black boxes, difficult to interpret and therefore ensuring physical consistency through such parameterizations is often challenging. Several examples of work have attempted to address this challenge. \cite{beuclerEnforcingAnalyticConstraints2021} apply nonlinear analytic constraints to a neural network as hard and soft constraints, reducing errors within the variables included in the constraints. \cite{heuerTrainingDataConfidenceGuided2025} reduced uncertainty for out-of-distribution input data by reverting back to the physics scheme through a confidence guided loss function, while distance based losses can constrain predictions to fall within the training data \citep{agarwalSkillfulGlobalOcean2026}. Further confidence and improvements can be gained by utilizing physically grounded parameterizations within the input feature space \citep{silvaPhysicallyRegularizedMachine2021}. Demonstrating why neural network closures behave as they do is a challenging research area, though interpretation of the what and why can be made if careful consideration is given to the data processing pipeline \citep{furtadoSettingStandardRecommended2026}.

The generalization of machine learning models to unseen regimes is a significant challenge to ensuring ML schemes reach adoption. Early work by \citet{ogormanUsingMachineLearning2018} revealed their subgrid ML closure for moist convection was unable to generalize to a warmer climate if the model was only trained on a control climate, with similar findings also found by \citet{raspDeepLearningRepresent2018}. \citet{boltonApplicationsDeepLearning2019} looked at how neural networks trained for one region generalize to other regions. They considered three distinct regions, and found that regions with the most turbulence generalized to other regions, likely due to the training region containing a diverse range of dynamics. \citet{bodnerDataDrivenApproachParameterizing2025a} showed that by holding out regions, their ML model could still perform well in that region, though whether performance remains when holding out multiple regions was not explored. \citet{perezhoginGeneralizableNeuralNetworkParameterization2025} designed a subgrid closure capable of generalizing across the resolution hierarchy. We do not attempt to build a generalizable subgrid closure, rather explore what could contribute to region to region generalization.

In this work we present developments of a machine learning mesoscale eddy parameterization based on the GEOMETRIC scheme \citep{Mak2018}. This work aims to take the first steps towards enhancing mesoscale process representation within the NEMO ocean model which serves as the ocean component for the UK's Earth System Model \citep{Sellar2019} at the eddy-permitting resolution. To achieve this, we explore in detail the performance of an offline machine learning model, where a future paper will explore its performance coupled to the NEMO ocean model. We are motivated in this work to explore model generalization - primarily owing to the computational costs in generating suitable training data. As such, we want to first answer some key questions: what aids model generalization to unseen regions; and how does the feature space impact model performance and generalization. We carry out this analysis using performance metrics, feature and target distributions, and spectral techniques. 

This paper is organised as follows. In Section \ref{sec:methods_theory}, we outline the methods used, from the physics based scheme to the training data, along with our choice in machine learning model. The results are then presented in three subsections: an overview of the regime and data distribution (Section \ref{ssec:regime}); model performance on held out regions (Section \ref{sec:holding-out-regions}); and a feature ablation study (Section \ref{sec:ablation-study}). We conclude with a discussion and summary.

%% file: methods-and-theory.tex
\section{Methods and theory} \label{sec:methods_theory}

\subsection{Scale-aware GEOMETRIC}

The scale-aware GEOMETRIC eddy parameterization scheme \citep{makScaleAwarenessEddyEnergy2023} is an adapted version of the standard GEOMETRIC parameterization \citep{Marshall2012, Mak2018}. The scale-awareness relates to the fact that the GM scheme only acts on scales larger than a length scale $L$, where traditional GM schemes acting at higher resolution damp eddies that are explicitly resolved \citep{Hallberg2013}. The scheme therefore adaptively predicts subgrid energy based on the horizontal resolution, where higher resolution grids such as $1/4^{\circ}$ and $1/12^{\circ}$ resolve more energy so require less parameterized eddy energy. The scale-aware GEOMETRIC equation is as follows,
\begin{equation}\label{eq:sa_GEOM}
	\partial_t \hat{E} + \nabla_H \cdot \left( \left( \tilde{\boldsymbol{u}}^z - |c|\boldsymbol{e}_x \right) \hat{E} \right) =
	\int \kappa_{gm}\frac{M^4_L}{N^2_L}dz - \lambda \left( \hat{E}-\hat{E}_0\right) +\eta_E \nabla^2_H \hat{E}
\end{equation}
where $\hat{E}$ is depth integrated eddy energy, $\tilde{\boldsymbol{u}}^z$ is the depth averaged flow, $|c|$ is the Rossby wave phase speed, $\kappa_{gm}$ is the GM coefficient, $M$ and $N$ are horizontal and vertical buoyancy frequencies, $\lambda$ is the dissipation rate, and $\eta_E$ is diffusion. The subscript $\cdot_L$ denotes a low-pass spatial filtering of scales smaller than $L$, which typically represents the separation scale between large and small scales. \citet{makScaleAwarenessEddyEnergy2023} set $L=100$ km, though this could be a varying scale over the global ocean. The energy tendency equation is solved using the current ocean state to predict eddy energy at the next time step. The variable $\hat{E}$ then informs the GM coefficient,
\begin{equation}\label{eq:sa_GEOM_kappa}
	\kappa_{gm} = \alpha \frac{\hat{E}}{\int M^2_L / N_L dz}
\end{equation}
which controls the parameterization for along isopycnal transport of tracers by mesoscale eddies. In Eq. \eqref{eq:sa_GEOM_kappa} the $\kappa_{gm}$ term also requires a tuning parameter, $\alpha$, of $O(10^{-1})$ and $\leq 1$.

By choosing to work with the scale-aware GEOMETRIC it offers an opportunity to leverage machine learning, while also maintaining a degree of confidence through the physics based term, $\kappa_{gm}$. This approach will enable us to mitigate many of the uncertain parameters in Eq. \eqref{eq:sa_GEOM} by bypassing the prognostic equation for $\hat{E}$. A number of approaches are available, such as the prediction of eddy energy tendency. In this case we focus on the prediction of instantaneous eddy kinetic energy ($KE_e$) for its ease of diagnosis. In fact, the form of \eqref{eq:sa_GEOM_kappa} may be less sensitive to the type of energy used \citep{bachmanEvaluationScalarEddy2017}. Furthermore, we predict only surface $KE_e$, where this field can be projected into the vertical through a structure function \citep{Ferreira2005}, surface baroclinic mode \citep{delalama2016}, or learned \citep{shanTestingLimitsSurface2026}. This projection however is not attempted in this current work but can be developed for the future. In addition, we also complement the work of \cite{parteeUsingMachineLearning2022} who predict eddy kinetic energy, though we make some adjustments to the overall process, which are discussed later.

\subsection{Numerical model}

In this study we utilise the Nucleus for European Modelling of the Ocean (NEMO) ocean model \citep{Madec2023}. We make use of NEMO to generate the training data, which we use to evaluate a machine learned prediction of subgrid eddy kinetic energy. 

The configuration we use in this study is the DIabatic Neverworld Ocean (DINO) \citep{kammDINODiabaticModel2025a}. The DINO model has been designed as both a means to generate training data and as a testbed for parameterizations, making it a natural choice for our study. The DINO configuration is based on the Neverworld2 model devised by \cite{marquesNeverWorld2IdealizedModel2022}, and resembles an idealized Atlantic Ocean sector covering both hemispheres. In the southern domain there is a periodic channel with a Drake Passage sill. The ocean is initialized from rest using simple temperature and salinity profiles (see Appendix D in \cite{kammDINODiabaticModel2025a}) and is forced at the surface with absolute wind stress and a seasonal solar heat flux. Both temperature and salinity are restored at the surface to ensure proper formation of high latitude dense water and maintenance of the thermal wind gradient for the Antarctic Circumpolar Current.

In contrast to the design in \cite{kammDINODiabaticModel2025a} we make some slight alterations. Instead of the Smagorinsky viscosity scheme, which is overly dissipative at the grid scale \citep{Bachmanetal2017}, we use a grid-aware biharmonic scheme. We also make the subjective choice to use the 4th order centered tracer advection scheme with explicit diffusion and use of a linear free surface. We use a horizontal resolution of $1/16^{\circ}$ to generate training, validation, and test data. We discuss the strategy to generate the training data in the following section.

\subsubsection{Training data}

The training data is generated from the high-resolution $1/16^{\circ}$ DINO configuration. The model is spun up for 60 years, which is adequate for the volume kinetic energy to stabilize. A further 12 years are simulated that are used for the machine learning model production phase. The data has a split of 10 years for training, and 1 year each for validation and testing. For the whole 12 years, each timestep is an instantaneous output. Training regions are chosen in an attempt to include various dynamical regimes for the machine learning model to learn from (see Section \ref{ssec:regime}). We choose a jet-like region (JET) that displays characteristic meandering and eddy generation, a downstream jet region (EAST) with weaker eddy activity, a southern ocean gyre region (GYRE) that mimics a high-latitude gyre regime, and finally a subtropical gyre (TROP) with very little activity. We anticipate that these four regions encompass a broad range of dynamics for the ML model to learn from. However, it is clear that some regimes not covered by the four regions might be omitted, such as the equatorial regions. We expect an eddy-permitting model to fully resolve mesoscale processes here, so a closure may not be required.

	\begin{figure}[t]
		\centering
			\includegraphics[width=0.6\textwidth]{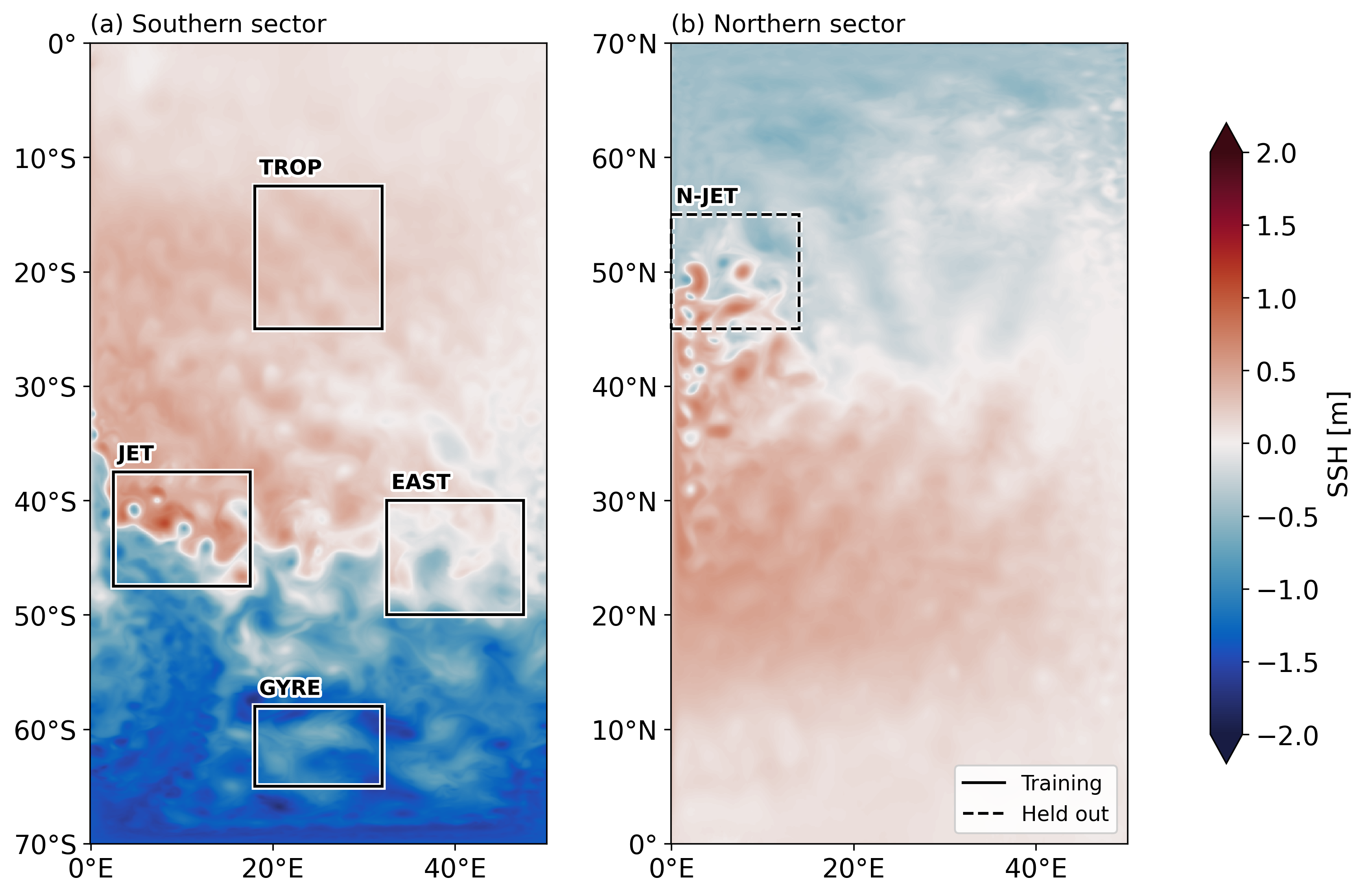}\\
  		\caption{DIabatic Neverworld Ocean (DINO), (a) southern hemisphere domain with training regions: JET (west), TROP (north), EAST (east), and GYRE (south), (b) northern hemisphere domain with held out test region, N-JET.}\label{fig:training_regions}
	\end{figure}

\subsection{Filtering and coarsening}

We now describe the filtering and coarsening approach that will enable us to train a machine learning model for our target resolution of $1/4^{\circ}$. The process of filtering and coarsening is a technique often used across machine learning for subgrid parameterizations (e.g. \citet{boltonApplicationsDeepLearning2019, bodnerDataDrivenApproachParameterizing2025a, jakharAnalyticalAIDiscoveredStable2026}). The process involves two-steps: typically applying a low-pass filter over high-resolution data; and then coarsening the data. These new fields then act as realizations of a lower resolution model, allowing a machine learned model to make predictions of subgrid energy. Several packages are available to filter spatial data \citep{storerFlowSieveCoarseGrainingUtility2023, nowakImplementationImplicitFilters2025}. Here we employ the Python package GCM-Filters \citep{looseGCMFiltersPythonPackage2022}. A filtered field is given by
\begin{equation}
	\overline{F}_l(\boldsymbol{x}) = G_l \ast F(\boldsymbol{x})
\end{equation}
where $G_l$ is a Gaussian filter that acts to remove scales smaller than $l$. Choices for $l$ vary between the input features and target. The function being filtered can take on the form of either a scalar or vector field. Following filtering, this field is then conservatively regridded from $1/16^{\circ}$ to $1/4^{\circ}$, which is carried out using the Python package, xesmf \citep{zhuangPangeodataXESMFV0922025}.

\subsection{Target and input features}

\subsubsection{Eddy kinetic energy}\label{sssec:eke}

One of the aims of this work is to predict instantaneous surface subgrid eddy kinetic energy ($KE_e$), which is our target variable, using a machine learning model. We compute eddy kinetic energy following \cite{buzzicottiSpatioTemporalCoarseGrainingDecomposition2023}
\begin{equation}
	KE_{e} = \frac{1}{2}\left(\overline{|\boldsymbol{u}|_l^2} - |\overline{\boldsymbol{u}}_l|^2\right)
\end{equation} 
where $\boldsymbol{u}=\left(u, v\right)$ are horizontal velocity components. In this case the filter length scale $l$ is chosen to be either of $\text{min}\left(3L_d, \Delta s\right)$ where
\begin{equation}\label{eq:BC1}
	L_d = \frac{1}{|f|\pi}\int_{-H}^0 N \, dz
\end{equation}
is the first baroclinic deformation radius, the length scale of mesoscale eddies, and $\Delta s$ is the average grid scale in the $1/4^{\circ}$ model. In each case, $3L_d$ or $\Delta s$ is the average in each region. Our purpose is to compute subgrid eddy kinetic energy, so through this choice of filter scale we are asking whether the eddy-permitting grid can resolve any $KE_e$, and if so, we only predict the subgrid component i.e. anything below the target grid (see table \ref{tab:filter-grid-scale}). This avoids any double counting that may occur if we chose a filter scale that ignores this. We choose mesoscale eddies to have length scales on the order of $3L_d$ \citep{Chelton2011}. In \eqref{eq:BC1} $N$ is the buoyancy frequency and $f$ is the Coriolis parameter. Eddy kinetic energy is then regridded to the target grid. We apply a natural log transform on eddy kinetic energy, as was done in \cite{parteeUsingMachineLearning2022}, meaning our target becomes $\ln\left(KE_e\right)$. Future work could look at diagnosing eddy available potential energy \citep{wenegratCoarsegrainedLocalAvailable2026}.

\begin{table}[h]
\caption{Filtering and grid scales}\label{tab:filter-grid-scale}
\begin{center}
\begin{tabular}{c c c c}
\hline\hline
Region & $\Delta s$ $1/16^{\circ}$ [km] & $\Delta s$ $1/4^{\circ}$ [km] & $3L_d$ [km] \\
\hline\hline
JET & $5.11$ & $20.41$ & $37.02$ \\
EAST & $4.9$ & $19.6$ & $31.23$ \\
TROP & $6.56$ & $26.26$ & $88.74$ \\
GYRE & $3.3$ & $13.19$ & $9.27$ \\ 
\hline\hline
\end{tabular}
\end{center}
\end{table}

\begin{figure}[t]
	\centering
		\includegraphics[width=1\textwidth]{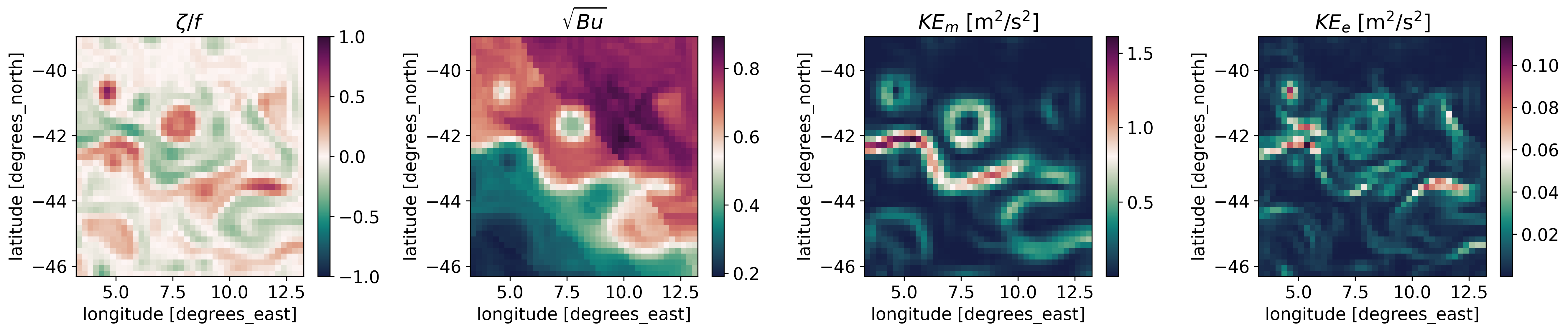}\\
  	\caption{Spatial maps of features and target in the JET region at a single time snapshot.}\label{fig:target_input_maps}
\end{figure}

\subsubsection{Features}

	To predict $KE_e$ we require input features that have a physical relation to the target. To this end we choose the following input features:
\begin{subequations}
	\begin{align}
		\zeta/f & = \left(\left(\nabla \times \boldsymbol{u}_{cg}\right)\cdot \boldsymbol{k}\right)/f \\
		KE_{m} & = \left(|\overline{\boldsymbol{u}_{cg}}_l|^2\right) \\
		\sqrt{Bu} & = \frac{\left(L_d\right)_{cg}}{\Delta s}
	\end{align}
\end{subequations}
where $\zeta/f$ is the normalized relative vorticity, $KE_{m}$ is coarse kinetic energy, and $Bu$ is the Burger number. The $\zeta/f$ feature can be thought of as an analog to the Rossby number and a measure of the importance of rotational effects, while the $\sqrt{Bu}$ is a measure of the length scale of mesoscale eddies with the grid-scale. Examples of these fields are shown in figure \ref{fig:target_input_maps}. The $\cdot_{cg}$ implies fields are filtered and coarsened. Velocities and tracers are filtered by removing scales below the $1/4^{\circ}$ grid, which are then used to compute the features. Coarse kinetic energy is further filtered after calculation to ensure it captures only scales above $3L_d$. Similarly to $KE_e$, we take a natural log transform of $KE_m$ as input into the machine learning model. Each feature being passed into the neural network is unitless after standardization. We could argue that our choice of feature space is generally invariant to the geographic location, so our choice of regions should generalize to the northern hemisphere. While this is a fairly simple feature space, additional features such as bottom topography may help the model to generalize further \citep{xieGlobalEstimationSubsurface2025}. In section \ref{sec:ablation-study}, we examine the importance of the feature space on model performance.

\subsection{Machine learning model for parameterization}

	Many approaches for subgrid ML closures employ a ML model based on a convolutional neural network \citep{boltonApplicationsDeepLearning2019}. The convolutional layers allow the model to capture spatial relationships between the features and target. This type of model will also allow us to make adaptations in future work, such as through the addition of temporal relationships. In this current work we use the U-Net model of \citet{ronnebergerUNetConvolutionalNetworks2015} to learn a functional mapping from the input features to the target. The U-Net model is a variant on the convolutional neural network, applying encoder and decoder blocks with skip connections. At each layer of the network, two convolution blocks are applied producing feature maps,  which are then downsampled. Further convolutions are applied until the bottleneck is reached. From there, the feature maps are upsampled. The distinguishing aspect of this model architecture are the skip connections, which pass and preserve fine spatial information that is lost during the encoder-decoder blocks. Variations of this model architecture have been applied as a surrogate for several Earth system model components, finding particular success in parameterizing sea-ice thickness \citep{durandDatadrivenSurrogateModeling2024} and atmospheric transport processes \citep{huStableMachineLearningParameterization2025}. The approach to apply the U-Net to a subgrid physics scheme is certainly underexplored and motivates its use in this study.
	
	To train the model we employ a mean squared error in the loss function and employ the ADAM optimizer with early stopping, where model weights are taken that provide the smallest error in loss. We apply replicative padding in the convolutional layers to preserve the input dimensions, which may aid performance at ocean boundaries \citep{zhangAddressingOutofSampleIssues2025}. We also normalize the input features using the Z-Score with data from the JET region. This choice is motivated in maintaining a single choice for all model training and ensuring mesoscale processes are normalized using data from an eddy rich region. Dimensions of the input and target are reduced to $40\times 40$ grid cells, removing any artifacts that may have come in from the data preparation process. Training is performed on the JASMIN ORCHID cluster on a Nvidia A100 GPU. Further training details are provided in Table \ref{tab:params_unet}.
	
\begin{table}[t]
\caption{U-Net model and training configuration}\label{tab:params_unet}
\begin{center}
\begin{tabular}{c c}
\hline\hline
U-Net details & \\
\hline\hline
Network depth & 2 \\
Base filters & 64 \\
Kernel size & $3 \times 3$ \\
Max pooling & $2 \times 2$ \\
Up-sampling & $2 \times 2$ \\
Activations & reLU, linear (final layer) \\
Optimizer & ADAM \\
Learning rate & 1e-3 \\
Early stopping & Measured from epoch 30 \\
Dropout regularization & 0.01 \\
Convolution padding & Replicative \\
Loss & Mean squared error \\
Batch size & 15 \\
\hline\hline
Data details & \\
\hline\hline
Split (train, val, test) & 0.83, 0.083, 0.083 \\
Training stride & 5 days \\
Training samples per region & 720 \\
Data dimension & $40 \times 40$ \\
Normalization & Zero mean, unit variance (JET-model) \\
\hline\hline
\end{tabular}
\end{center}
\end{table}

%% file: results.tex
\section{Results}

\input{regime-analysis.tex}

\input{holding-out-regions.tex}

\input{holding-out-features.tex}

%% file: regime-analysis.tex
\subsection{A brief exploration of the data}\label{ssec:regime}

	We first examine the feature space ($\zeta/f$, $KE_m$, $\sqrt{Bu}$) and their relation with the target ($KE_e$) across all regions. This will inform our U-Net based model development and deliver optimal predictive capabilities.  
	
\begin{figure}[h]
		\centering
			\includegraphics[width=0.9\textwidth]{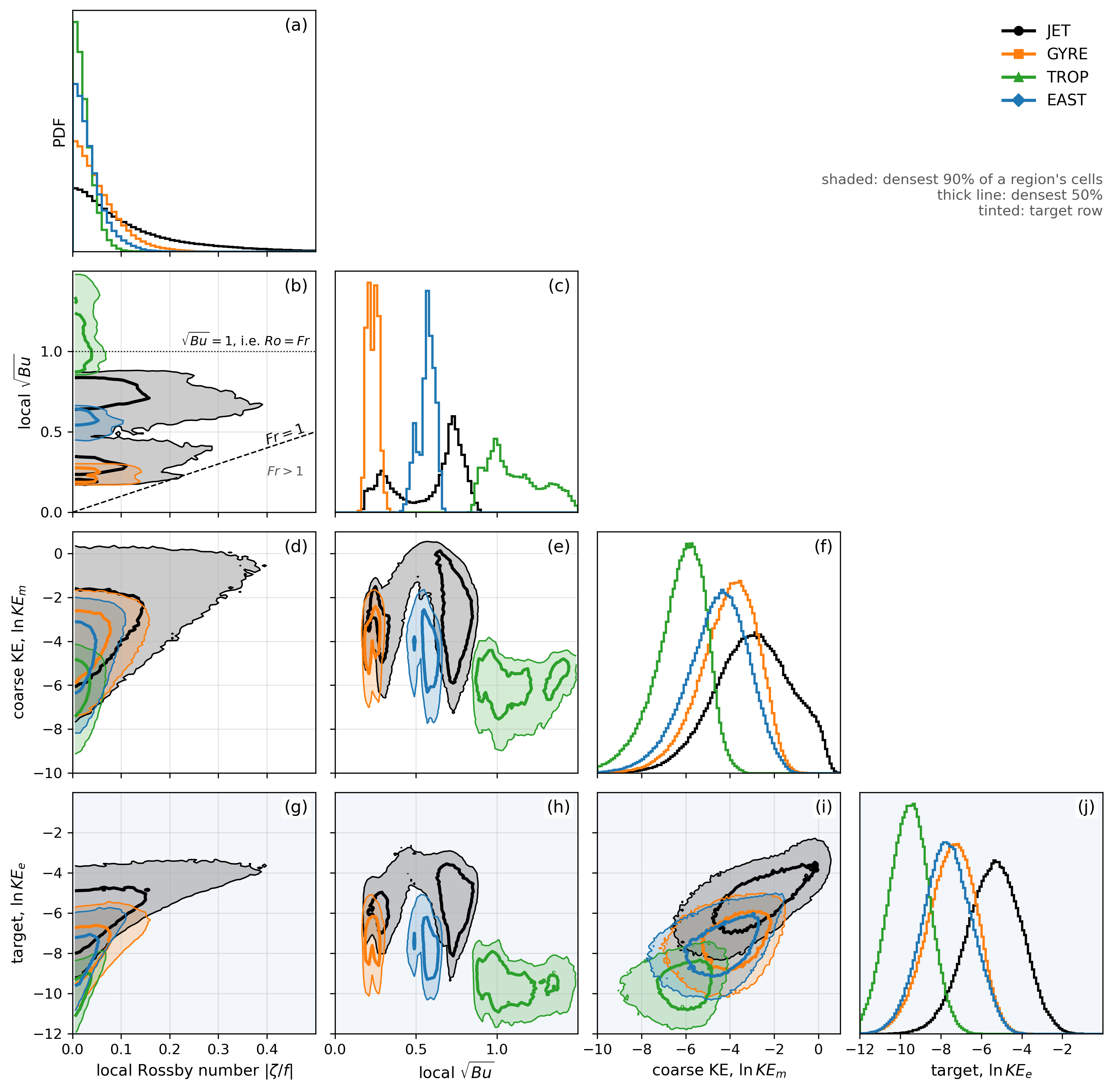}\\
  		\caption{The relationship of input and target distributions. Probability density functions are shown on the diagonal for the input features (a,c,f) and target (j). Two-dimensional feature space relationships in (b,d,e), and feature vs target relationships in (g,h,i).}\label{fig:regime-features}
	\end{figure}

	Figure \ref{fig:regime-features} displays the distributions of the inputs and the target in 2D space along with their probability density functions. The data shown here is from the test set, however, it also adequately describes the training set as  there was no significant difference between the test and training set distributions. Figure \ref{fig:regime-features}(b) shows a regime space by plotting $|\zeta/f|$ against $\sqrt{Bu}$, along with dotted lines for $Fr=1$ and $Ro=Fr$. The $Fr$ is the Froude number, which is the ratio $Ro/\sqrt{Bu}$. When $Ro < Fr$ or $\sqrt{Bu} < 1$, the regime is dominated by rotation, but when $Ro > Fr$, this is a regime of both rotation and stratification. From this we see that the only region outside of the rotation dominated regime is TROP, as a result of its weak Rossby number and large Burger number. We see that the training data sits in a regime of $Bu\in\left(0.02,2\right)$, where only TROP and the north-east component of JET sit closest to the Qausi-Geostrophic limit, $Bu\approx 1$, while GYRE and the south-west component of the JET region display the smallest $Bu$. The JET distribution displays a double structure, where the region with the small Burger number corresponds to the shallow sill in the Drake Passage. These latter regions demonstrate poorly resolved mesoscale regions. The range of $Bu$ here illustrates partly the problem with the capacity of eddy-permitting models to represent the mesoscale \citep{Hallberg2013}, where part of the domain can resolve eddies but parts cannot, and hence the need for resolution adaptive parameterizations. The Rossby number space hints at a mostly geostrophic limit with $|\zeta/f|\ll 1$, though some weakly ageostrophic motions may exist in the JET region where we expect strong fronts to accompany the eddy motions. This also suggests that the $KE_e$ prediction may include some ageostrophic motions, which is not entirely desirable for a parameterization of the balanced flow. 
	
	Figures \ref{fig:regime-features}(d,e) represent the feature space between $ln\left(KE_m\right)$ against $|\zeta/f|$ and $\sqrt{Bu}$, respectively. In the $\left(ln\left(KE_m\right),\sqrt{Bu}\right)$ space, the TROP region is separated by its $\sqrt{Bu}$, while its densest 50\% of $ln \left(KE_m\right)$ and $|\zeta/f|$ are captured by the JET. We might expect that models trained on anything but TROP will struggle to extrapolate to high energy regions, owing to the PDF tails in figure \ref{fig:regime-features}(a,c,f,i). We also see that JET seems to capture much of the spread in feature space, with exceptions in TROP with weak $lnKE_m$ and large $\sqrt{Bu}$ (figure \ref{fig:regime-features}(d,e)), as well as in the $\left(ln\left(KE_m\right),\sqrt{Bu}\right)$ space in EAST. We note that the $\zeta/f$ feature can display some skewness in observations and realistic simulations \citep{shcherbinaStatisticsVerticalVorticity2013}, but due to our idealized setup here, we do not see a significant skew in either direction (not shown). The target $ln \left(KE_e\right)$ is also related to each individual feature (figure \ref{fig:regime-features}(g,h,i)). Here we see that the distribution of $ln \left(KE_e\right)$ is similar to $ln \left(KE_m\right)$, where they are positively related. 
	
\begin{figure}[h]
		\centering
			\includegraphics[width=1\textwidth]{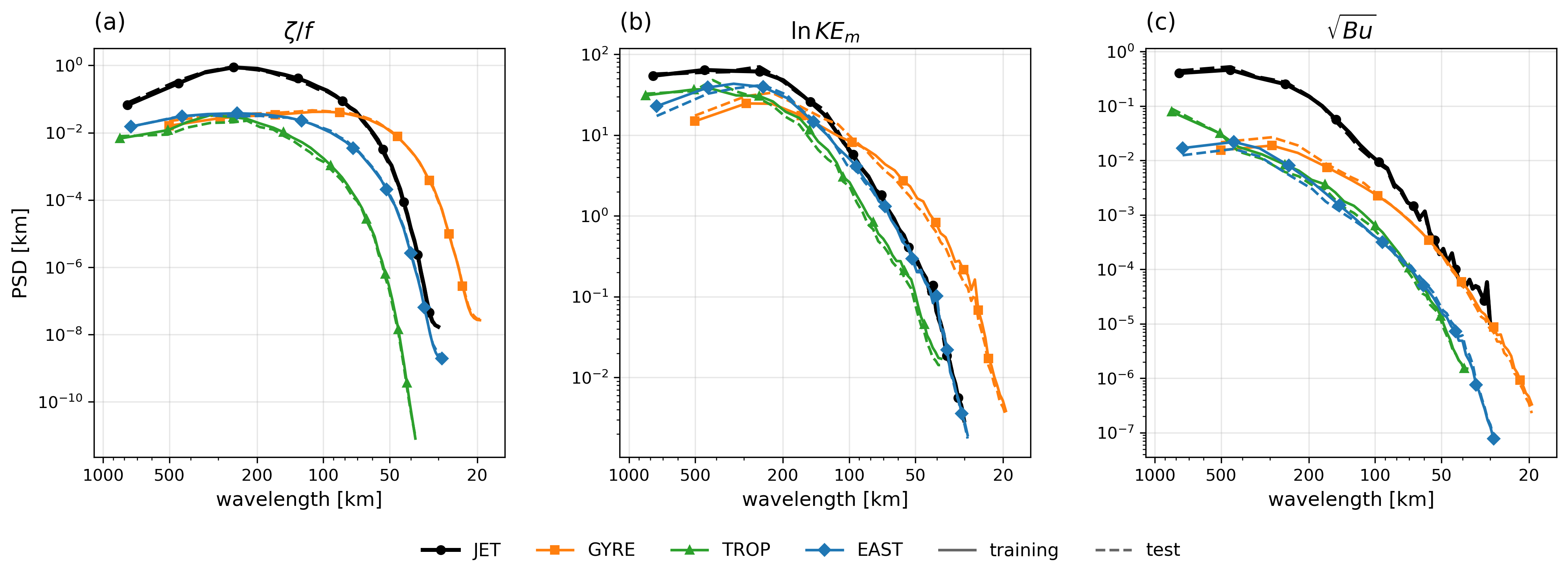}\\
  		\caption{Power spectral density (PSD) of the input features: (a) $\zeta/f$, (b) $ln \left(KE_m\right)$, and (c) $\sqrt{Bu}$. The PSD is computed at each snapshot then time-averaged. Each region is colored and full lines are the training set and dashed lines are the test set. Input data is unitless. Data is first regridded to a cartesian equidistant grid. The PSD is computed using the xrft Python package \citep{uchidaXgcmXrftV1012023}.}\label{fig:psd-features}
	\end{figure}

	Although we see some overlap of input feature space across regions, a power spectral density (PSD) analysis of the feature space can help reveal whether the regions share similar spectral properties. This may also further inform how well a trained model can generalize across regions. Figure \ref{fig:psd-features} displays the PSD for each feature in each region, for both the training and test data set. In both $\zeta/f$ and $ln \left(KE_m\right)$ (figure \ref{fig:psd-features}(a,b)) we see higher latitude regions capture PSD at smaller wavelengths, shifting their peak magnitude from larger to smaller wavelengths as well, particularly in $\zeta/f$. This shift in latitude is due to changes in grid spacing (see table \ref{tab:filter-grid-scale}). We also see in $\zeta/f$ and $ln \left(KE_m\right)$ that the JET and EAST regions share similar spectra at small wavelengths, but EAST displays weaker PSD at larger wavelengths due to it being downstream of the stronger JET region. Both the TROP and GYRE regions are distinct from each other, a result of the opposing latitudes (low vs high). The $\sqrt{Bu}$ has a slightly different spectra (figure \ref{fig:psd-features}(c)), with TROP and EAST sharing a similar slope, and JET and GYRE sharing a similar slope at small wavelengths. The difference in $\sqrt{Bu}$ is likely owing to the inclusion of the grid spacing.
	
%% file: holding-out-regions.tex
\subsection{Model performance and region generalization}\label{sec:holding-out-regions}

	We now present results from the evaluation of the machine learning prediction of subgrid eddy kinetic energy. In this section we assess how the machine learning model performs by training it on combinations of the four regions, which results in 15 individual models. A key question we want to answer is regarding drivers and barriers for generalization to  unseen regions. This will help us formulate the necessary modeling framework for online parameterization schemes (integrated or coupled with NEMO) in the future. The analysis begins with an evaluation of the model scores, which are used as a reference point for model generalization. We show sensitivity findings in appendix \ref{ssec:appendix-sen-model-params}.
	
	\begin{figure}[h]
		\centering
			\includegraphics[width=1\textwidth]{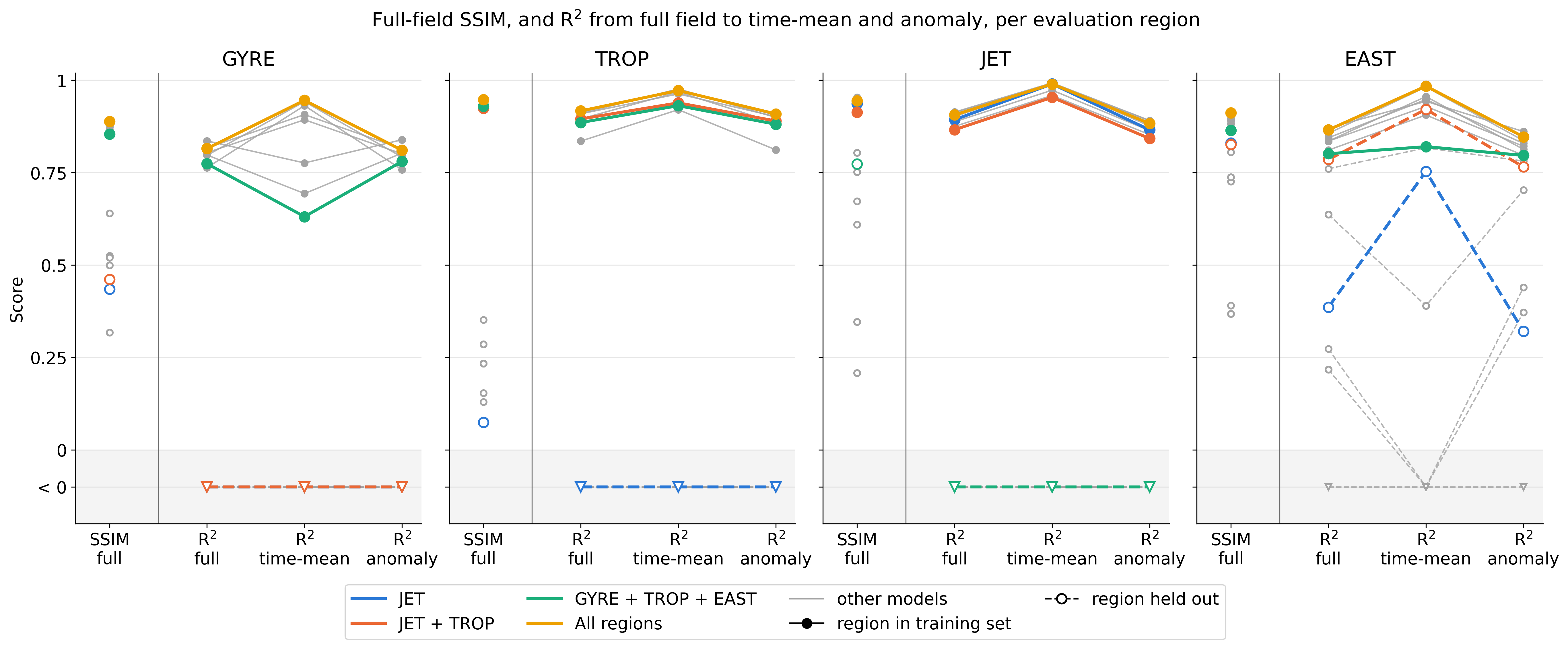}\\
  		\caption{Model R2 (full, time-mean, anomaly) and SSIM scores in each region when holding out regions for four select models: JET, JET-TROP, GYRE-TROP-EAST, All regions. Regions that are in the training data are given by the full lines, whereas regions outside of training are indicated by the dashed lines. Negative R2 scores are capped at zero, shown by the model lines with small triangles. Scores are calculated using the non-transformed field, $KE_e$.}\label{fig:regional_model_scores_regions}
	\end{figure}
	
	The R2 and SSIM scores enable us to benchmark the performance of the trained ML models, where we focus on the following: JET, JET-TROP, GYRE-TROP-EAST, and All regions (figure \ref{fig:regional_model_scores_regions}). A heatmap showing all model scores can be seen in figure \ref{ssec:full-scores}. We also present the decomposed R2 score into its time-mean and anomaly components, which will further show how well the model predicts the mean state and temporal pointwise deviations. From the score metrics in figure \ref{fig:regional_model_scores_regions} we see that the model trained on All regions is one of the best performing models, predicting accurately ($R2\approx 0.95$) the time-mean state in all regions, however, all possess R2 anomaly scores less than their time-mean R2. In this model and many others, the time-mean R2 is larger than its R2 full and anomaly, which is consistent with previous studies that show data-driven closures tend to capture the average statistics well \citep{rossBenchmarkingMachineLearning2023} but smooth out the small scale features \citep{subichFixingDoublePenalty2025}. We do see the opposite of this in other models (e.g. GYRE-TROP-EAST-model), but we believe this is a result of a prediction bias in the mean state. The R2 scores are generally consistent over time (figure \ref{fig:n-jet-held-out}(a)), though the performance in GYRE exhibits some poor performing time-periods.
	
\begin{figure}[h]
		\centering
			\includegraphics[width=1\textwidth]{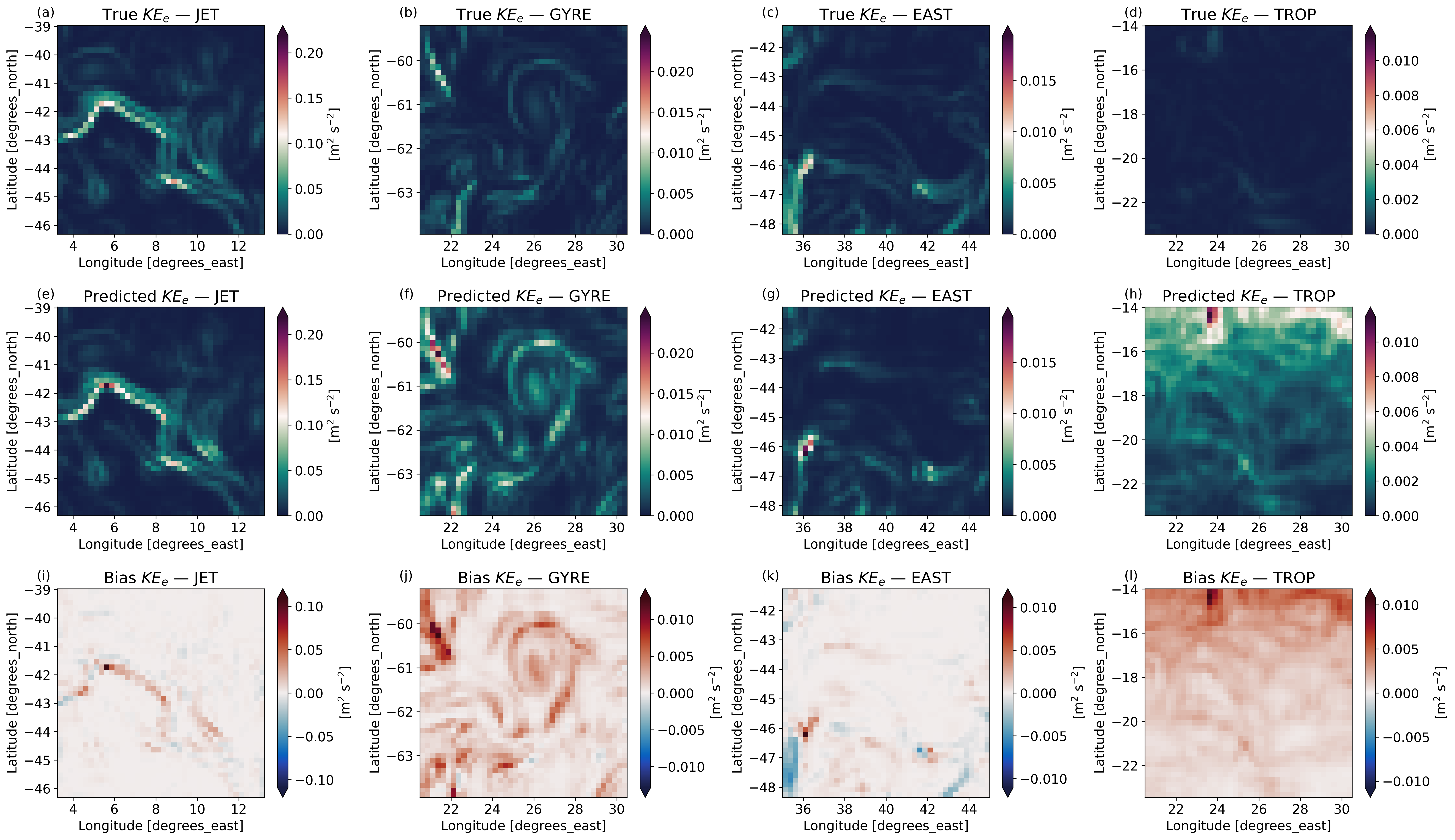}\\
  		\caption{Instantaneous snapshot of subgrid $KE_e$ over each region with true fields in (a) JET, (b) GYRE, (c) EAST, and (d) TROP. Predictions are based on the JET-model (e,f,g,h) along with their respective bias (i,j,k,l).}\label{fig:eke-maps-true-pred-bias-jet}
	\end{figure}

	We now focus on the JET-model in figure \ref{fig:regional_model_scores_regions} where it performs well in its own region with comparable performance to the model trained on All regions. The JET-model performs poorly in GYRE and TROP regions, with only a modest SSIM score in the GYRE region, suggesting at least some structure is captured by the JET-model. Recall from section \ref{ssec:regime} that the TROP region displays very little data overlap with the JET region, potentially explaining the poor performance of the JET-model in that region. In the GYRE region, there is more data overlap with the JET region, to a similar extent to the EAST region, but generalization doesn't occur. The performance in the EAST region is promising, and shows some generalization capability, with an SSIM score of $0.83$ and a time-mean R2 score of $0.75$. However, the full and anomaly R2 score sit at $0.3-0.4$, implying the JET-model does not capture the variance well. Context to these scores are shown in figure \ref{fig:eke-maps-true-pred-bias-jet} through instantaneous snapshots of $KE_e$ along with its bias. All bias plots show varying levels of pointwise error, with particular hotspots of overprediction (figure \ref{fig:eke-maps-true-pred-bias-jet}(i,j,k,l)). The JET region has quite large errors across its jet core (up to $0.1$ m$^2$s$^{-2}$), with some smoothing contributing to those errors. In the GYRE region, the magnitude of the error is smaller but mostly positive, consistent with the low R2 and modest SSIM scores. In the EAST region, the bias exhibits fewer hotspots, with concentration in the South of the domain. The positive bias in the TROP region is indicative of the poor model scores and lack of generalization. \citet{boltonApplicationsDeepLearning2019} found western boundary regions were most successful at generalizing to other regions, consistent with our findings here.
	
	We find that including both the JET and TROP regions in the training data leads to some improvements, notably in the EAST region (full R2 rising from $0.39-0.79$), scoring as well as the GYRE-TROP-EAST-model. Within the TROP region lie some of the distribution tails in the EAST region (figure \ref{fig:regime-features}(d,i)), which could contribute to its improved performance. In models that do not include the JET region, such as the GYRE-TROP-EAST-model, we find they struggle to reproduce the $KE_e$ field in the JET region. By just examining the distribution (figure \ref{fig:regime-features}), its clear to see that none of the other regions match the intense values of the JET region, as such the model does not know how to predict JET. This is a distributional shift, where both the input and target distributions shift, and could be a proxy for how the model may perform out-of-distribution e.g. warmer climate. This also implies that if we want a model to perform well in a particular region, then it must be trained with this data.
	
\begin{figure}[h]
		\centering
			\includegraphics[width=0.9\textwidth]{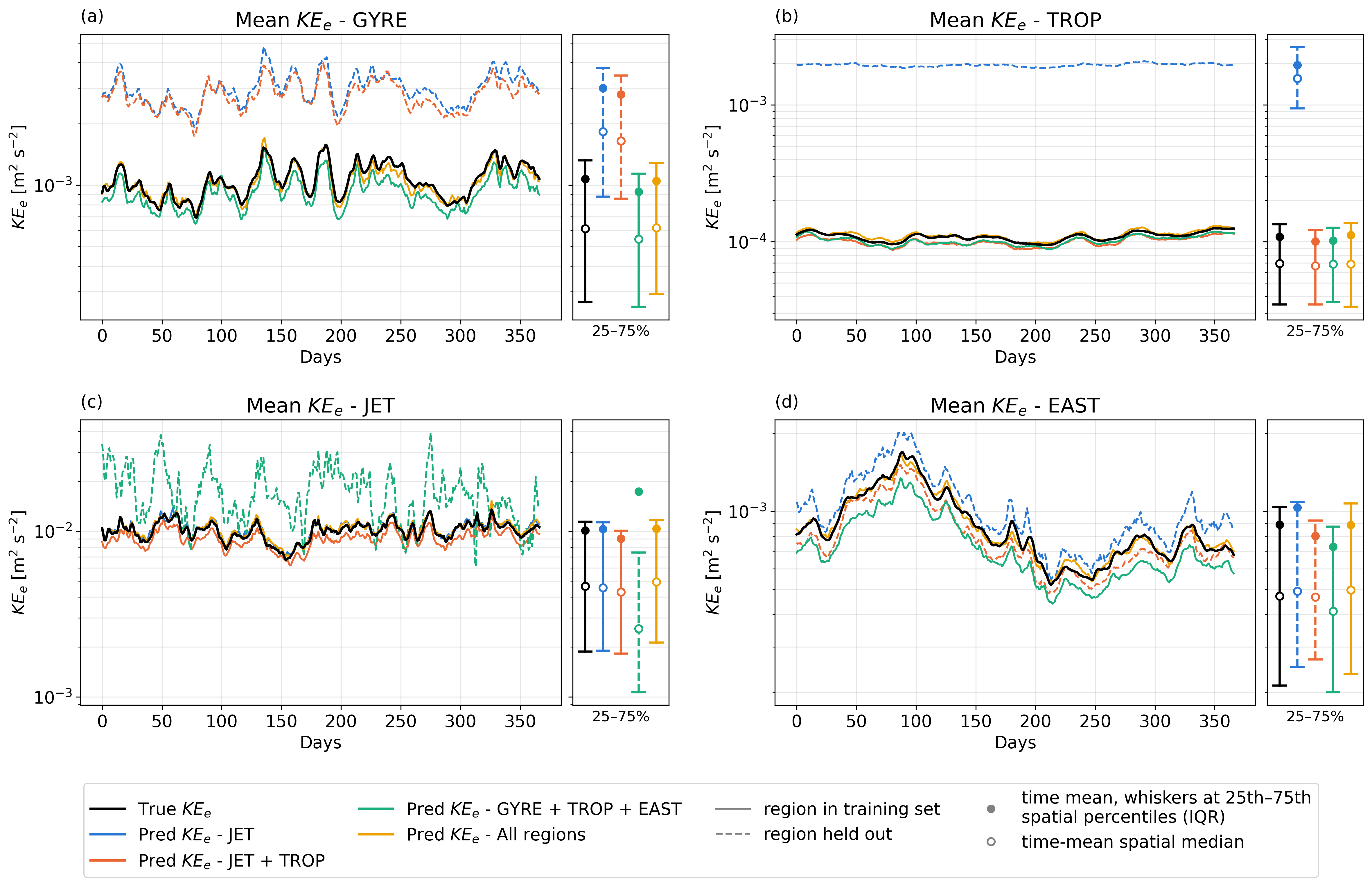}\\
  		\caption{Area weighted mean $KE_e$ for four models, JET, JET-TROP, GYRE-TROP-EAST, All regions, over the test set in each region: (a) GYRE, (b) TROP, (c) JET, (d) EAST. Side panels show the upper 75th and lower 25th quartile range, with median (clear dot) and mean (shaded dot). Full lines are regions in the training set, dashed lines are regions outside the training set.}\label{fig:regional_mean_eke_regions}
	\end{figure}

	The model scores provide an overview of the models performance, but they say nothing about how the predictions vary over time or how they are distributed. A time-series of domain averaged $KE_e$ is shown in figure \ref{fig:regional_mean_eke_regions} for the same four models: JET; JET-TROP; GYRE-TROP-EAST; All regions. It's clear from many of the models that they are able to capture the mean state variability well, even so for the JET- and JET-TROP- models in the GYRE region (figure \ref{fig:regional_mean_eke_regions}(a)) despite their poor model scores. Exceptions to this are the JET-model in the TROP region and the GYRE-TROP-EAST-model in the JET region. We see that the poor scores for the JET- and JET-TROP- models in the GYRE region are down to a positive bias, consistent with the instantaneous snapshots in figure \ref{fig:eke-maps-true-pred-bias-jet}(b,f,j). The overprediction in the GYRE region may not be a distribution problem, but in fact tied to the filtering length scale that we used. This could be related to the $KE_e$ spatial average filter scale, where JET contains two distinct modes in $\sqrt{Bu}$ (figure \ref{fig:regime-features}(c)). It could also be due to the fact that the model has intrinsically learned to predict $KE_e$ below the eddy-permitting grid, which is the case for the regions JET, TROP, and EAST. However, in the GYRE region, $KE_e$ is defined at scales further below the grid (see table \ref{tab:filter-grid-scale}). Without this training information, the model is unable to extrapolate to the GYRE region, and subsequently overpredicts $KE_e$. In fact, if we adjust the filter scale for the calculation of $KE_e$ in GYRE from $3L_d$ to $6L_d$ ($9-18$ km), we find the JET-model is able to capture the true field, though not for the correct reasons (see figure \ref{fig:gyre-filter-scale}).
	
	Alongside each panel in figure \ref{fig:regional_mean_eke_regions} are bars displaying the inter-quartile range, and mean and median. These show generally that regions in the training set are consistent with the truth. We've shown that the JET-model generalizes to the EAST region, and does so well over a seasonal cycle that is not present in other regions. In the model without the JET region (GYRE-TROP-EAST-model) the prediction of the JET region is poor, where we see an underprediction of the median and inter-quartile range but an overprediction of the time-mean. This further indicates the importance of including a range of data distributions in training (e.g. \citep{ogormanUsingMachineLearning2018, raspDeepLearningRepresent2018}, or to consider methods to improve out-of-distribution performance \citep{saynisch-wagnerConditionalUpdatesNeural2026}.

\subsubsection{Fully held out region - a northern boundary JET}

\begin{figure}[h]
		\centering
			\includegraphics[width=0.9\textwidth]{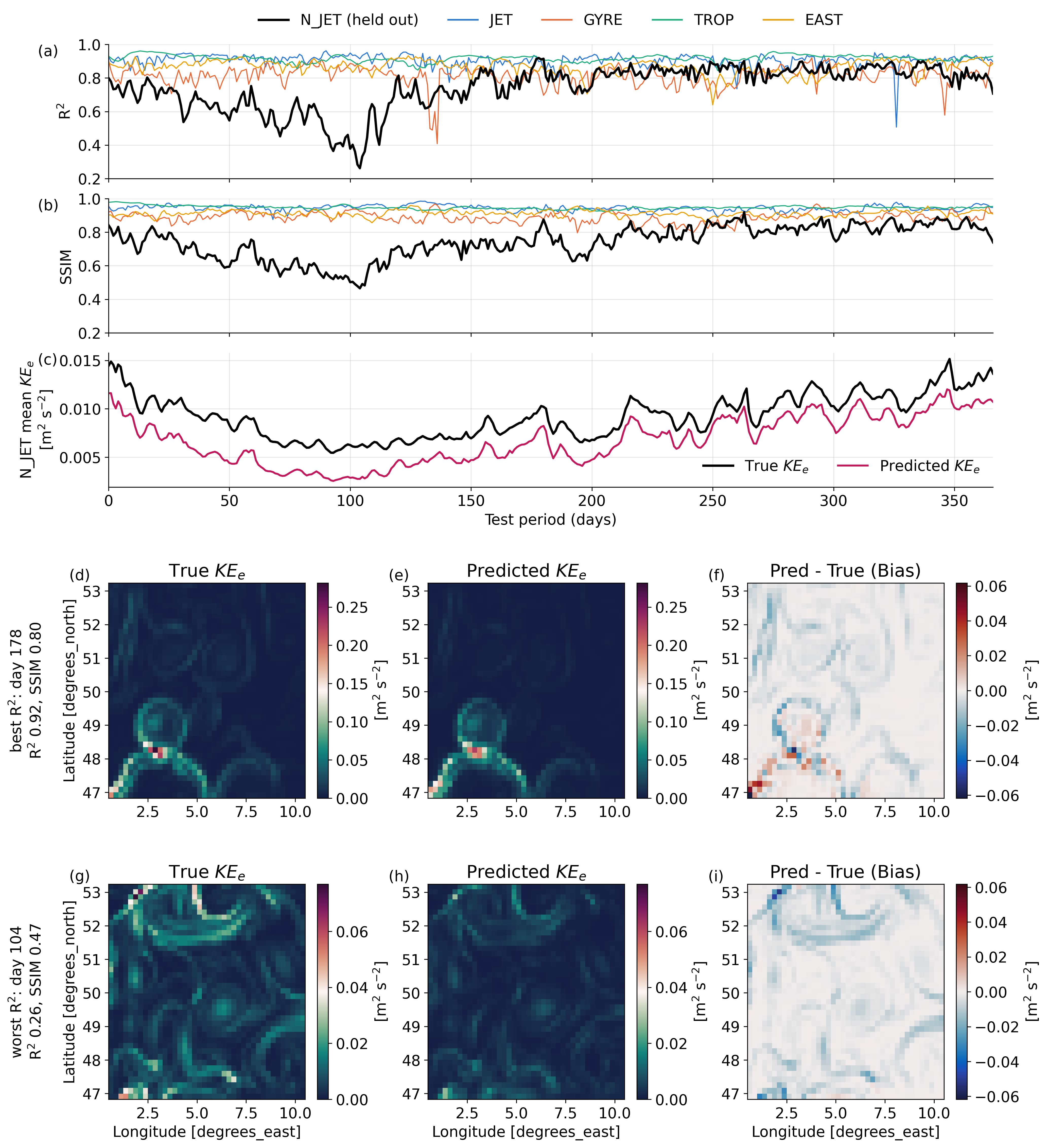}\\
  		\caption{Performance of the All regions all features model in the fully held out region, N-JET. In (a) daily full R2 score, (b) daily SSIM score, (c) time-series of area-weighted $KE_e$. Spatial maps in best (second bottom row) performing day with (d) true $KE_e$, (e) pred $KE_e$, and (f) pred-true, and worst performing day (bottom row) with (g) true $KE_e$, (h) pred $KE_e$, and (i) pred-true.}\label{fig:n-jet-held-out}
	\end{figure}

	We now test the All regions and All features model on a fully held out region. We choose the N-JET region, which sits in the northern sector and hugs the coastline. This will enable us to validate whether a southern sector trained model can generalize to a region in the northern hemisphere. Figure \ref{fig:n-jet-held-out} shows time-series of model scores and $KE_e$, along with spatial maps of $KE_e$ at the best and worst performing days. Generally, the R2 and SSIM are good. For the first 100 days of the test period, the R2 and SSIM scores drop, before recovering from day 150 (figure \ref{fig:n-jet-held-out}(a,b)). It can be seen in both the best and worst predictions (figure \ref{fig:n-jet-held-out}(e,h)) that some of the mesoscale activity is underpredicted (figure \ref{fig:n-jet-held-out}(f,i)). In the best performing day, the model is able to capture a significant feature to the South of the domain, while on the worst day, the model fails to capture the true magnitude. We can see in the mean $KE_e$ time-series in figure \ref{fig:n-jet-held-out}(c) that the worst period coincides with weak $KE_e$, while the improved performance has $KE_e$ levels almost two times stronger. There could be a number of reasons for the poor extrapolation to N-JET during the initial period when the scores drop off: lack of seasonal variation in JET; weak seasonal variation in EAST; latitude and thus $\sqrt{Bu}$ difference.
	
%% file: holding-out-features.tex
\subsection{Dependence on input feature space}\label{sec:ablation-study}

\begin{figure}[h]
		\centering
			\includegraphics[width=1\textwidth]{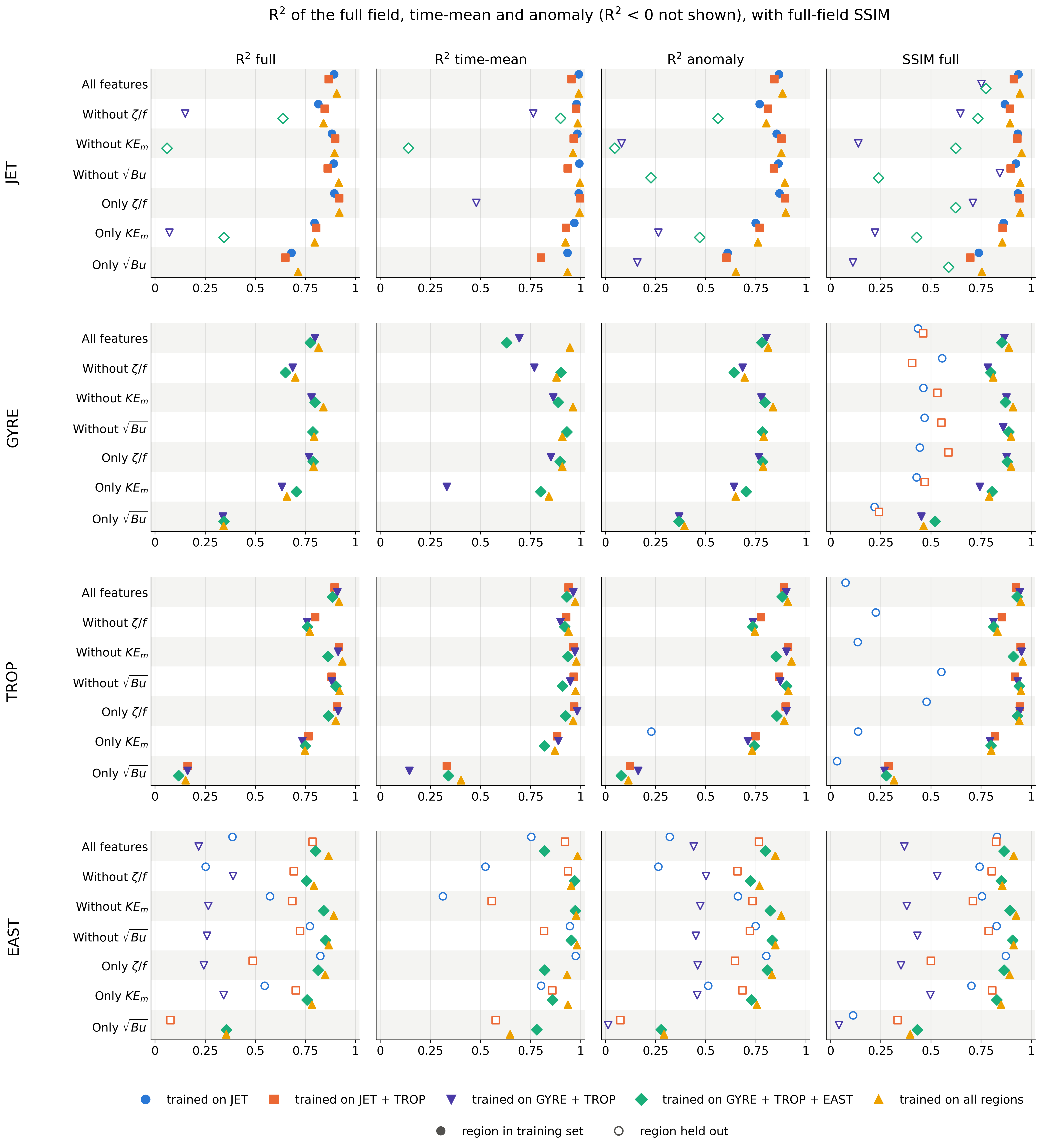}\\
  		\caption{Model R2 (full, time-mean, anomaly) and SSIM scores in each region when holding out features for five select models: JET, JET-TROP, GYRE-TROP, GYRE-TROP-EAST, All regions. Regions that are in the training data are given by the shaded shape, whereas regions outside of training are indicated by the clear shape. Negative R2 scores are left off.}\label{fig:feature_ablation_heatmap}
	\end{figure}

	We now present a feature ablation experiment to determine how changes in the feature space influence model performance and whether we can offer any explanations for any changes seen. For each region defined model, we re-train them each a further six times across the combination of two and one variable feature spaces. Figure \ref{fig:feature_ablation_heatmap} displays the full and decomposed R2 and SSIM score metrics of four specific models: JET; JET-TROP; GYRE-TROP; GYRE-TROP-EAST; and All regions. This gives us a broad overview of how model performance changes depending on its feature space.
	
	We begin with the JET-model, where we recall in section \ref{sec:holding-out-regions} that the model could reproduce the mean climatology in the EAST region. Overall, we find that holding out features does not lead to any shifts towards region generalization, despite minor increases in SSIM in the GYRE and TROP regions. There are some minor reductions in R2 scores in the JET region when using the single feature spaces  only-$\sqrt{Bu}$ and only-$KE_m$. This reduction mostly happens in the R2 anomaly, suggesting these feature spaces degrade the models ability to capture the true variance. In the EAST region, we see shifts in model performance across feature spaces. Several models show a significant increase, with full R2 moving from $0.39$ in All features to $\sim 0.8$ for without-$\sqrt{Bu}$ and only-$\zeta/f$. Here, both of these improvements can be deduced from figure \ref{fig:regime-features}(e,g), with $\sqrt{Bu}$ sitting in the JET gap and JET's $\zeta/f$ mapping fully to EAST's $lnKE_e$. Without-$KE_m$ also sees an increase in full R2, but this comes from the R2 anomaly, while the time-mean R2 decreases. In the EAST region, it appears that the inclusion of $\sqrt{Bu}$ degrades the performance. The explanation for the poor only-$\sqrt{Bu}$ scores is simply because no mapping of this feature to the target exists in the JET training data (figure \ref{fig:regime-features}(h)).

	In the JET-TROP-model, the picture is not too dissimilar to the JET-model, with good performance in each region, JET and TROP. In the GYRE region, we do not see any changes in region generalization when changing the feature space, despite distributions between GYRE and EAST being similar e.g. $KE_m$. In the TROP region, ablation performance is generally consistent with the JET region, though performance is very poor when using only-$\sqrt{Bu}$. In the EAST region, the changes across feature ablations display a smaller range, with performance in All features greatly improved from the JET-model. Interestingly, the only-$\zeta/f$ model exhibits a poor R2 time-mean and decrease in R2 anomaly, contrary to the JET-model. We attribute this to a model underprediction (not shown). We nevertheless see all scores decrease from the All features model, which could suggest over a multi regime space, we may need multiple features for accurate predictions. 
	
	In the GYRE-TROP-model, there is an almost random drop in performance when holding out $\sqrt{Bu}$, which we identify as an edge artifact, potentially a result of the replicative padding (see appendix \ref{ssec:appendix-model-blow-up}). 
	
	We saw poor performance of the GYRE-TROP-EAST-model in the JET region when using All features. Here, we begin to see hints of region generalization to the JET region in the feature ablations. Traces of this may already be shown through the good SSIM score, implying the model is able to reproduce the correct structure. Removing $\zeta/f$ results in a full R2 score of $0.64$, however this is the only model able to generalize here. This is in contrast to the JET-model in the EAST region where $\zeta/f$ aided model performance. This may suggest that different regions may require different feature spaces, or some feature weighting, to achieve the best performance.
	
\begin{figure}[h]
		\centering
			\includegraphics[width=0.9\textwidth]{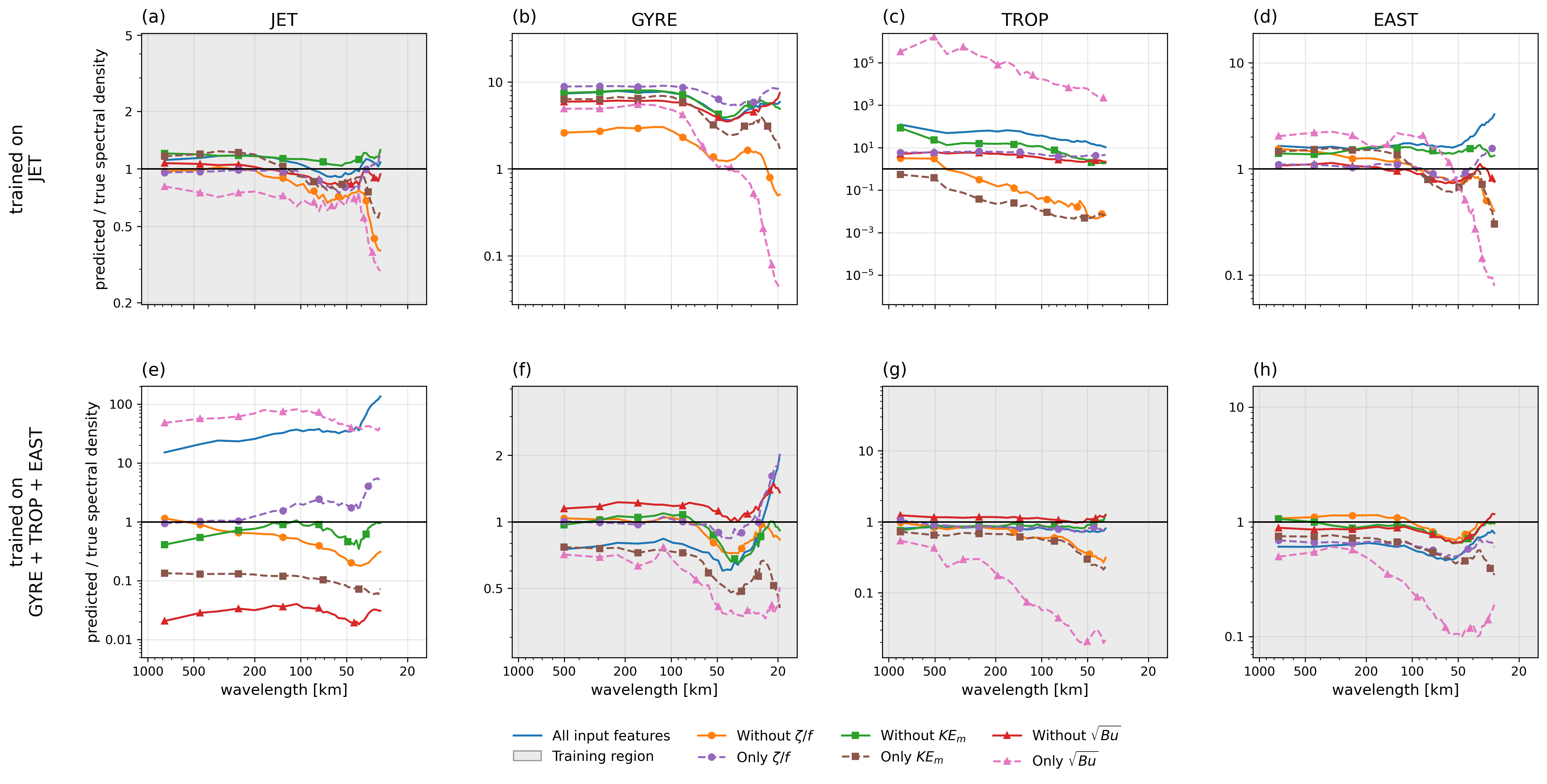}\\
  		\caption{The ratio of the power spectral density between the predicted and true $KE_e$ in each region, (a,e) JET, (b,f) GYRE, (c,g) TROP, (d,h) EAST, for two models, (top) JET-model, (bottom) GYRE-TROP-EAST-model. The PSD is computed at each snapshot then time-averaged. Data is first regridded to a cartesian grid. The PSD is computed using the xrft Python package \citep{uchidaXgcmXrftV1012023}.}\label{fig:KEe-psd}
	\end{figure}

	A power spectral density of the ratio between predicted and true $KE_e$ now reveals how the feature space modulates the predictions across wavelengths and reveals finer details than the model scores (figure \ref{fig:KEe-psd}). Here we discuss only two models: JET, and GYRE-TROP-EAST. We omit the previous models for clarity. In the regions that display good model performance (shaded panel and EAST in JET-model), we see in most a divergence from unity at smaller scales, indicating a degradation in prediction. This is a well known occurrence in ML models that employ the mean squared error loss function, a result of spectral bias \citep{chattopadhyayLongtermInstabilitiesDeep2023, balwadaDesignImplementationDataDriven2026}. In the JET-model in the GYRE region (figure \ref{fig:KEe-psd}(b)) we notice that there is a constant offset ($500-70$ km), implying the JET-model is capturing a similar slope, but overpredicts the magnitude. In the JET-model's own training region (figure \ref{fig:KEe-psd}(a)), the only-$\zeta/f$ feature space could be the best model due to its close alignment of spectra over most wavelengths, while the all features JET-model shows greater spectra at larger scales. By removing $KE_m$, the spectra is in better agreement at the smallest scales. In the GYRE-TROP-EAST-model in the JET region, the only-$KE_m$ feature space has an almost constant offset from larger scales down to about $75$ km. When dropping $\zeta/f$, the ratio is around 1 for larger scales, but drops at smaller scales, while using only-$\zeta/f$ has good agreement at large scales but the prediction error grows at smaller scales. These spectra are complementary to model scores, showing that poor scores might not necessarily display poor spectral similarity.
	
%% file: conclusion.tex
\section{Summary and discussion}

	Machine learning is offering the ocean modeling community new tools to tackle long-standing problems within climate modeling. As such, the aim of this work was to demonstrate the first steps in adapting a physics based mesoscale eddy parameterization scheme using machine learning for eventual integration with the UK's Earth System Model. The parameterization in question was chosen to be the scale-aware GEOMETRIC scheme \citep{makScaleAwarenessEddyEnergy2023}, building on the body of research developing Gent-McWilliams type ML-informed schemes \citep{parteeUsingMachineLearning2022,balwadaDesignImplementationDataDriven2026}. This allowed us to take the relatively straightforward approach of predicting instantaneous surface subgrid eddy kinetic energy from a physical input space using a well established neural network based tool/algorithm, the U-Net. From this we conducted a detailed offline analysis by comparing the neural network predictions with the true $KE_e$ field.
	
	The U-Net was trained on three input features, $\zeta/f$, $\sqrt{Bu}$, and $KE_m$, which kept the model lightweight and easier to interpret. We applied a similar feature space to \cite{parteeUsingMachineLearning2022}, though we omitted the mean isopycnal slope and instead normalized relative vorticity by the Coriolis parameter. We did explore how these features were related to each other and the target, which helped us gain a prior insight into how the models trained on individual selected regions might generalize to unseen regions and what impact feature ablations could have on model performance. We found our complete dataset to sit in a predominantly rotation dominated regime ($Ro<Fr$), with limits close to quasi-geostrophy and geostrophy ($Ro<1$ and $Bu\approx 1$), typical of an eddy-permitting model.
	
	Findings from the offline machine learning model show overall good skill in predicting subgrid $KE_e$. Using all regions and all features, the model achieved time-mean R2 scores of $R2\approx 0.95$, with full field R2 scores of $0.26-0.9$ in a fully held out North Atlantic JET region. We found some models were able to generalize to unseen regions, such as the JET- and JET-TROP- models for the EAST region. However, this was not the case for other regions e.g. GYRE, with reasons pointing to the filter scale used in computing $KE_e$ from the $1/16^{\circ}$ model. Further non-generalization (e.g. in GYRE-TROP-EAST-model) was found to be a result of distribution shift, with an unseen input and output space, indicating limitations of the ML models capabilities. We found some shift towards generalization when adjusting the feature space. For example, the JET-models performance in the EAST region improved when using only-$\zeta/f$, which produced $KE_e$ spectra more aligned with the true field, suggesting the feature $\zeta/f$ may carry much of the variability to reproduce $KE_e$. Though in some cases, model generalization was found when removing $\zeta/f$ e.g. in GYRE-TROP-EAST-model in the JET region. These results suggest there is a complex interplay between feature space and where the trained models are deployed. This also suggests adaptations in features or filtering may be necessary before applying the scheme into NEMO.
    
    Despite earlier work by \citet{ogormanUsingMachineLearning2018, raspDeepLearningRepresent2018} who showed that extrapolation beyond training data could be difficult, or impossible, which we indeed find to be the case, we nevertheless conducted this exploration as a necessary and low-cost first step. Recent developments show that generalization across resolutions can be achieved by adapting the input feature space using dimensional analysis \citep{perezhoginGeneralizableNeuralNetworkParameterization2025, balwadaDesignImplementationDataDriven2026}, whilst generalization to out-of-distribution data (e.g. increase in stratification) is realized by modifying weights in a neural network \citep{saynisch-wagnerConditionalUpdatesNeural2026}, or through transfer learning \citep{darmanFourierAnalysisPhysics2026}. Moreover, architectural design could also play a part in generalization \citep{shanTestingLimitsSurface2026, rautPerspectiveSelfSupervisedTransformers2026, ghoshPhysicsembeddedTransformerCNNArchitecture2026}. Adapting the methodology to predict a tendency term may also be more suitable \citep{chenNeuralOrdinaryDifferential2019} enabling the scheme to embed memory from previous ocean states. Many of these developments could enhance our current parameterization and should be explored. In addition, it may also be worth asking whether a targeted regional subgrid closure may be more relevant, such as in regions where we know models struggle to resolve mesoscale processes e.g. high-latitude regions.
	
	Here, we performed an offline supervised learning approach. Since the eventual closure is not being plugged into a dynamic equation (e.g. \citep{maulikSubgridModellingTwodimensional2019}), we might hypothesize that this will cause fewer problems when coupled online. However, it could result in  physical model drift in temperature and salinity patterns as the ML model will likely need re-tuning for the new scheme \citep{perezhoginGeneralizableNeuralNetworkParameterization2025}. Alternative approaches to closure development lie in online learning, where the closure is learned directly alongside the target model, typically through calibration or trajectory learning. This approach generally requires the numerical solver to be differentiable, which is not the case for typical Fortran/C++ based models \citep{frezatPosterioriLearningQuasiGeostrophic2022, oualaOnlineCalibrationDeep2024}. Applying online learning to the GEOMETRIC equations could come in the form of learning the parameters in the prognostic equation, such as eddy dissipation \citep{guanPredictionExtremeEvents2026}, or alternatively via a purpose built Earth system model that is machine learning ready \citep{gentineLegoESMModularDifferentiable2026}.
	
	We have evaluated the machine learned prediction of subgrid eddy kinetic energy using a relatively simple set of diagnostics. By also focusing on a small subset of regions from the DINO configuration, we have been able to explore whether a ML model will perform well or not. As such, this work lays the groundwork for later deployment of the $KE_e$ prediction into the GEOMETRIC closure and evaluation in NEMO and UKESM runs, with further work focusing on online stability and global performance. Future work should look at evaluating how this closure performs in realistic simulations, particularly focusing on regions that are crucial for long-term climate stability.

%% file: appendix-A.tex
\section{Appendices}

\subsection{Sensitivity to filter scale}\label{ssec:appendix-sen-filter-scale}

	Increasing the filter scale for the calculation of $KE_e$ in GYRE results in a closer match between the JET-model's GYRE prediction. This illustrates the deficiency in this region, suggesting adaptations to the feature space or architecture may be needed to improve predictions where the $KE_e$ filter scale sits below the target $\Delta s$.

	\begin{figure}[h]
		\centering
			\includegraphics[width=1\textwidth]{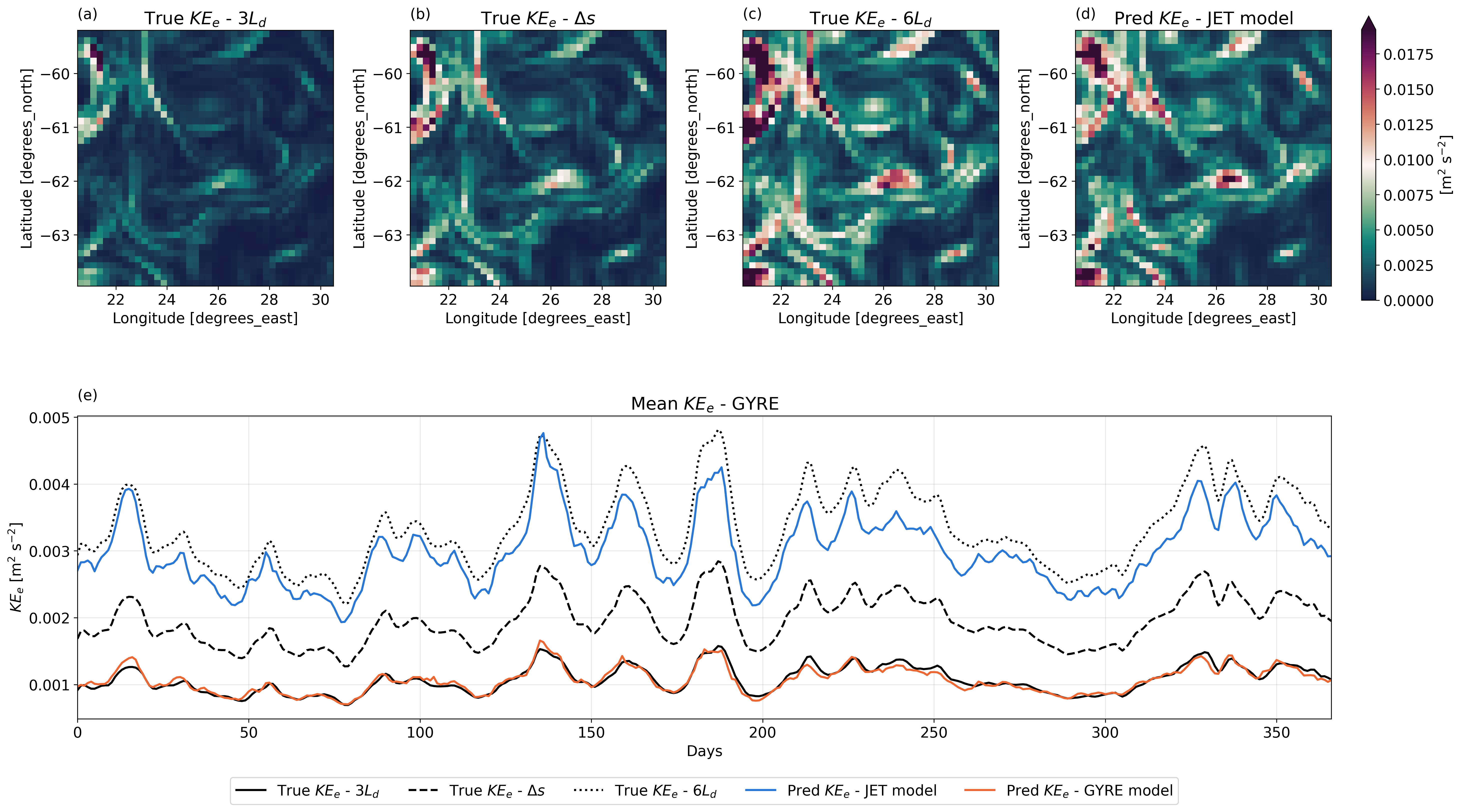}\\
  		\caption{In (a,b,c) true $KE_e$ fields of an instantaneous snapshot in the GYRE region computed using different filter scales: $3L_d$, $\Delta s$, $6L_d$, against the JET trained GYRE region prediction (d). Time-series in (e), which includes the prediction from the GYRE-model.}\label{fig:gyre-filter-scale}
	\end{figure}
	
\subsection{Sensitivity to model and training parameters}\label{ssec:appendix-sen-model-params}

	\begin{figure}[h]
		\centering
			\includegraphics[width=0.7\textwidth]{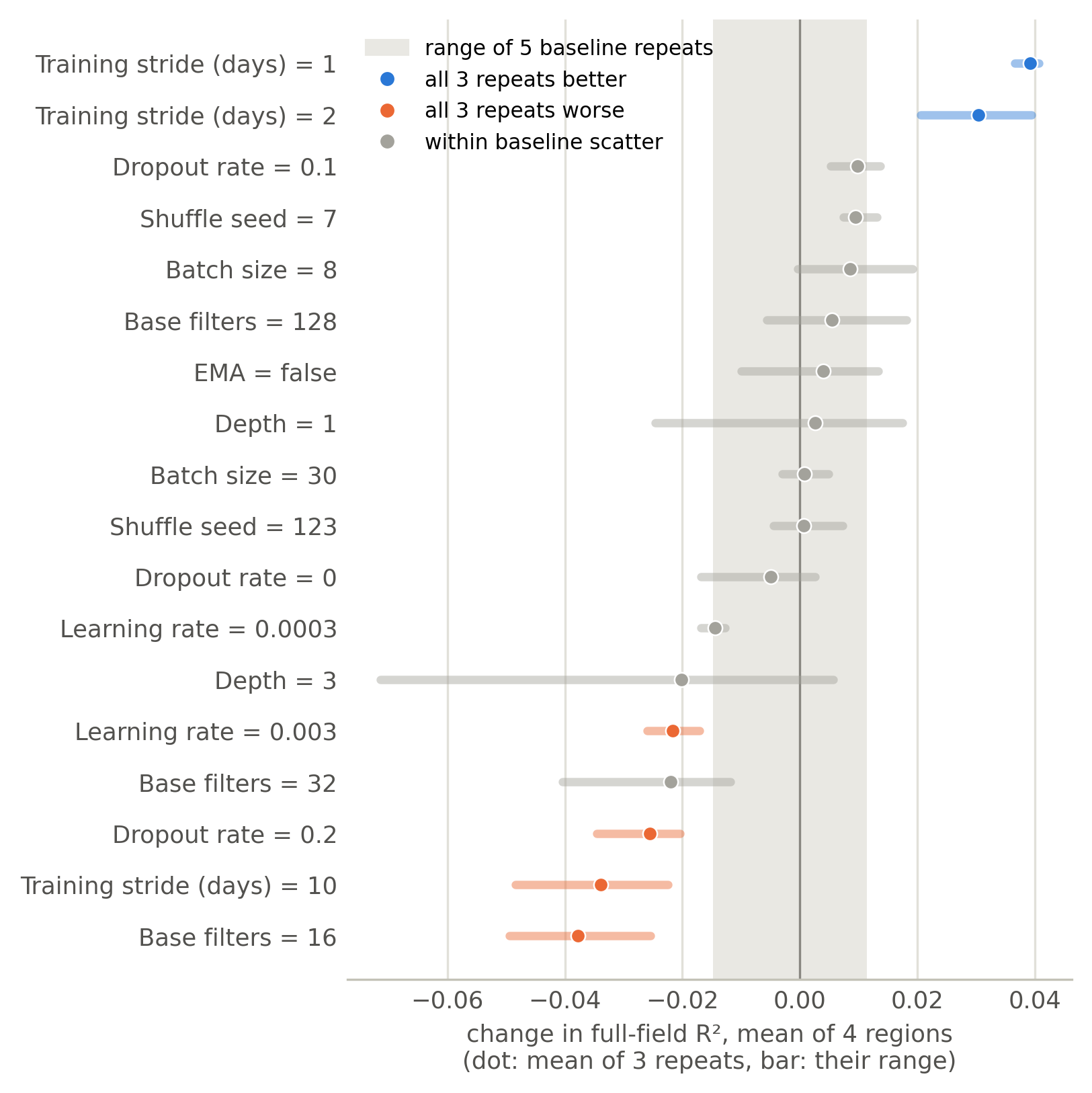}\\
  		\caption{Changes to the full-field R2 score in the all regions all features model from sensitivity sweep over model and training parameters.}\label{fig:sensitivity-tests}
	\end{figure}
	
	\clearpage
	
\subsection{Illustrating poor GYRE-TROP-model ablation performance}\label{ssec:appendix-model-blow-up}

	The model blow up in the GYRE-TROP-model in the GYRE region when holding out $\sqrt{Bu}$ is an edge artefact. When the prediction window is shifted to the left, the prediction error disappears.

	\begin{figure}[h]
		\centering
			\includegraphics[width=0.7\textwidth]{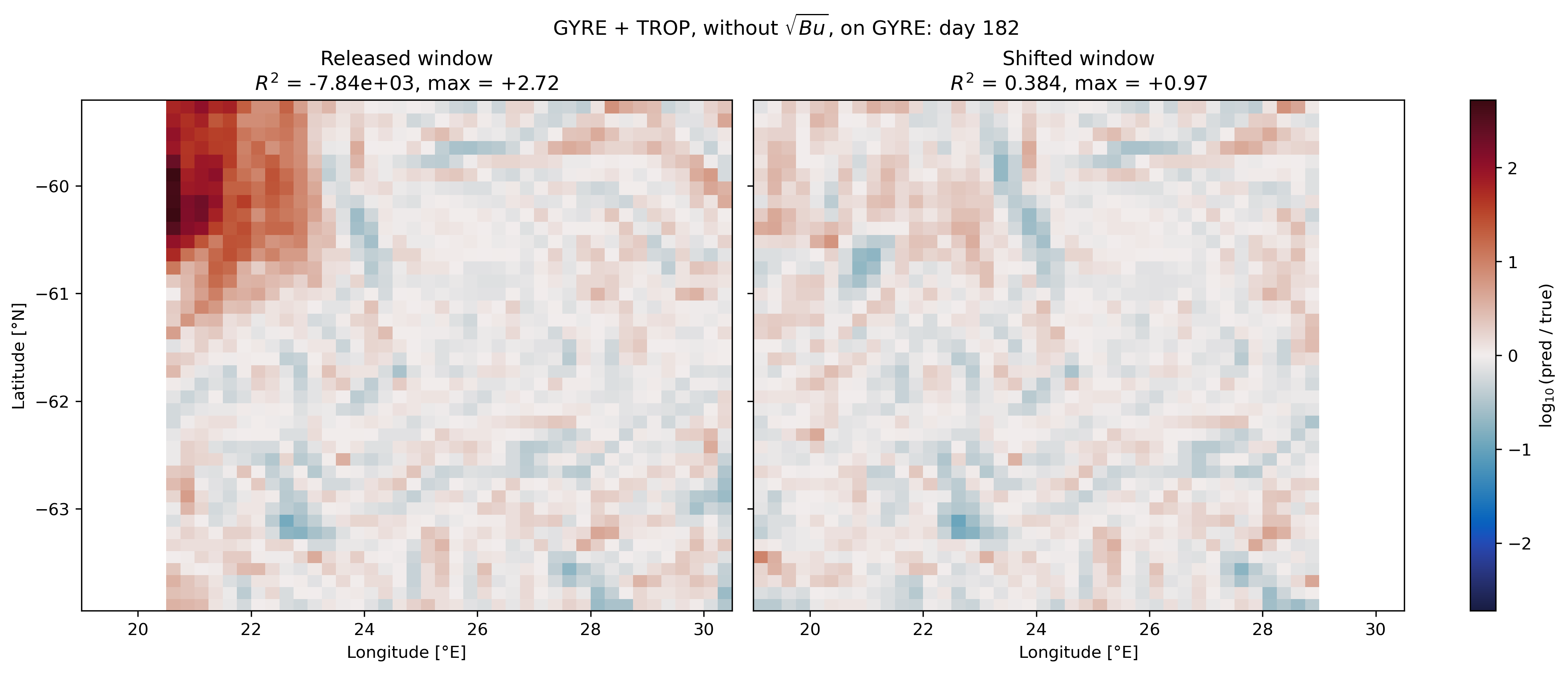}\\
  		\caption{$log_{10}\left(pred/true\right)$ of the GYRE-TROP-model holding out $\sqrt{Bu}$ in the GYRE region, showing the effect of shifting the prediction window on the model blow.}\label{fig:gyre-trop-blow-up}
	\end{figure}
	
	\clearpage
	
\subsection{Scoring metrics}

\subsubsection{Coefficient of determination}

	To measure the accuracy of the prediction we use the coefficient of determination, or R-squared (R2), given by,

	\begin{equation}\label{eq:r2-score}
		R^2 = 1-\frac{\sum^N_{n=1}\left(y_n-\hat{y}_n\right)^2}{\sum^N_{n=1}\left(y_n-\overline{y}_n\right)^2}
	\end{equation}
	
	which measures the proportion of variance accounted for by the model. In the above, $y$ is the true field and $\overline{y}$ its mean, and $\hat{y}$ is the predicted field. For an accurate prediction, we expect values to be closer to $1$, while negative values indicate poor predictions.

\subsubsection{Structual Similarity Index Measure}

	In addition to R-squared, we use the structural similarity index measure (SSIM)
	
	\begin{equation}
		SSIM\left(\hat{y},y\right) = \frac{\left(2\mu_{\hat{y}} \mu_y + C_1\right)\left(2\sigma_{\hat{y},y}+C_2\right)}{\left(\mu^2_{\hat{y}}+\mu_{y}^2+C_1\right)\left(\sigma^2_{\hat{y}}+\sigma^2_y+C_2\right)}
	\end{equation}
	
	which measures the spatial consistency between the predicted and true fields. For similarly structured fields, we expect values to be close to 1, while poor predictions close to or equal to $0$.
	
	\clearpage
	
\subsection{Full R2 and SSIM scores}\label{ssec:full-scores}

	\begin{figure}[h]
		\centering
			\includegraphics[width=1\textwidth]{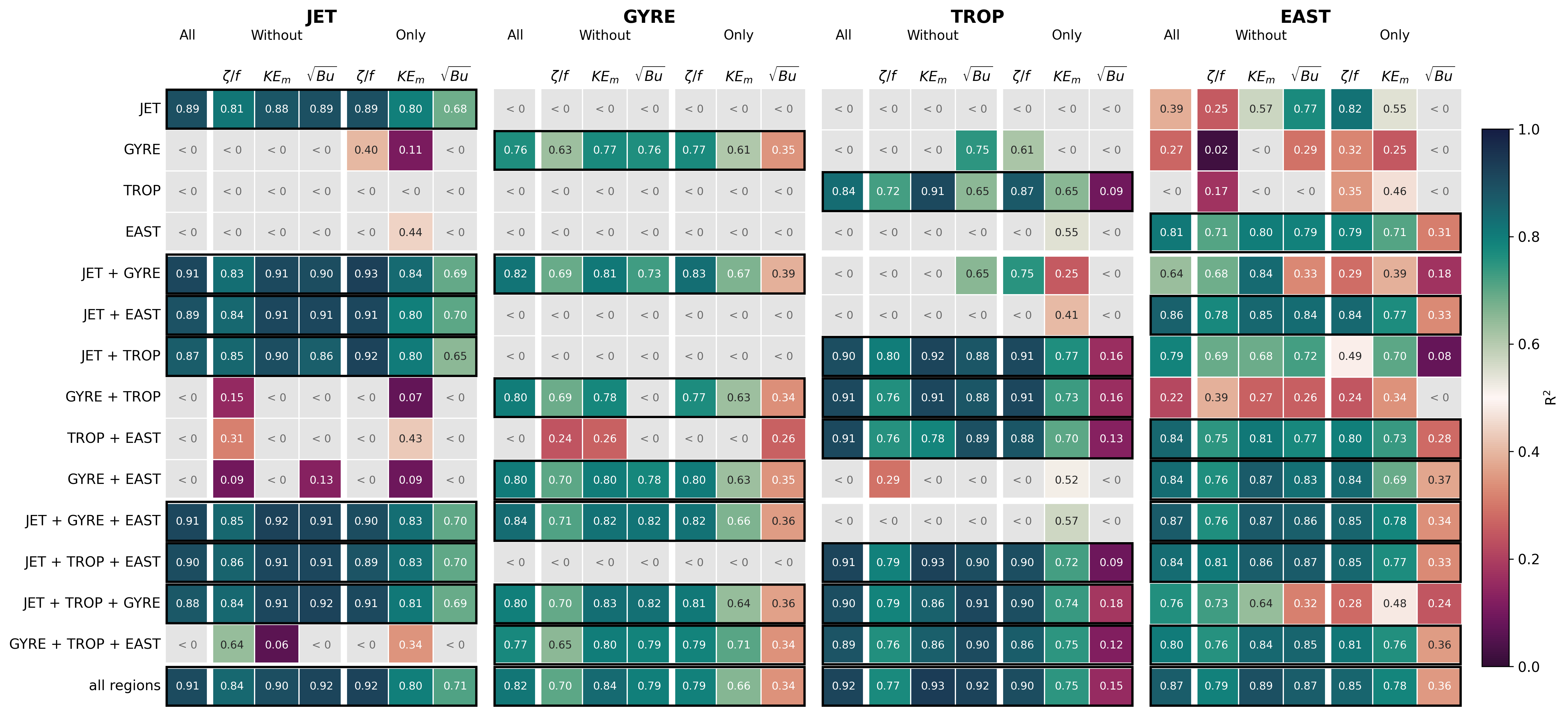}\\
  		\caption{The full-field R2 score for all models, regions, and feature ablations.}\label{fig:r2-scores-all}
	\end{figure}
	
	\begin{figure}[h]
		\centering
			\includegraphics[width=1\textwidth]{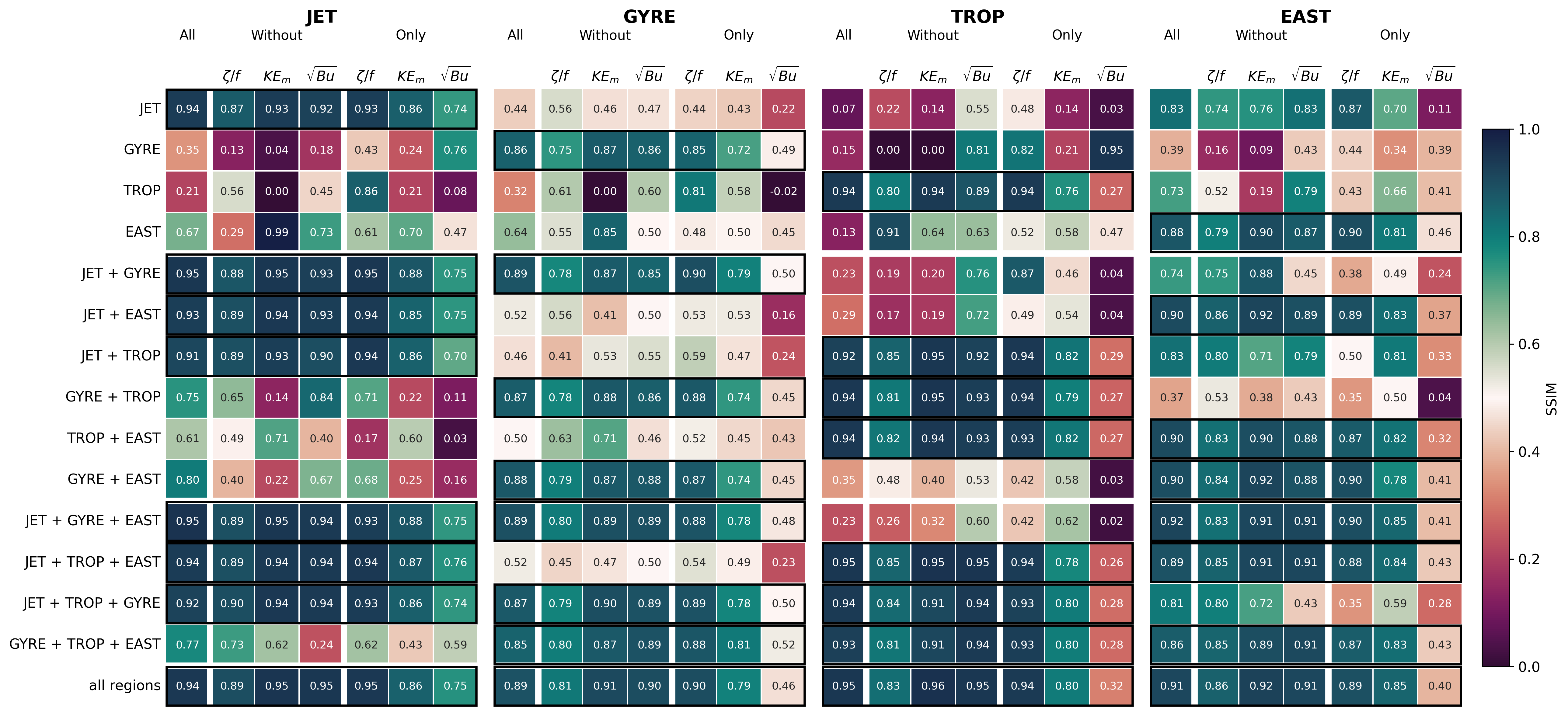}\\
  		\caption{The full-field SSIM score for all models, regions, and feature ablations.}\label{fig:ssim-scores-all}
	\end{figure}

%% file: main.bbl
\begin{thebibliography}{76}
\providecommand{\natexlab}[1]{#1}
\providecommand{\url}[1]{\texttt{#1}}
\expandafter\ifx\csname urlstyle\endcsname\relax
  \providecommand{\doi}[1]{doi: #1}\else
  \providecommand{\doi}{doi: \begingroup \urlstyle{rm}\Url}\fi

\bibitem[Agarwal et~al.(2026)Agarwal, Smith, Frolov, and
  Slivinski]{agarwalSkillfulGlobalOcean2026}
N.~Agarwal, T.~A. Smith, S.~Frolov, and L.~C. Slivinski.
\newblock Skillful {{Global Ocean Emulation}} and the {{Role}} of
  {{Correlation-Aware Loss}}, Apr. 2026.

\bibitem[Armour et~al.(2024)Armour, Proistosescu, Dong, Hahn,
  {Blanchard-Wrigglesworth}, Pauling, Jnglin~Wills, Andrews, Stuecker,
  {Po-Chedley}, Mitevski, Forster, and
  Gregory]{armourSeasurfaceTemperaturePattern2024}
K.~C. Armour, C.~Proistosescu, Y.~Dong, L.~C. Hahn,
  E.~{Blanchard-Wrigglesworth}, A.~G. Pauling, R.~C. Jnglin~Wills, T.~Andrews,
  M.~F. Stuecker, S.~{Po-Chedley}, I.~Mitevski, P.~M. Forster, and J.~M.
  Gregory.
\newblock Sea-surface temperature pattern effects have slowed global warming
  and biased warming-based constraints on climate sensitivity.
\newblock \emph{Proceedings of the National Academy of Sciences}, 121\penalty0
  (12):\penalty0 e2312093121, Mar. 2024.
\newblock \doi{10.1073/pnas.2312093121}.

\bibitem[Bachman(2019)]{Bachman2019}
S.~D. Bachman.
\newblock The {{GM}}+{{E}} closure: {{A}} framework for coupling backscatter
  with the {{Gent}} and {{McWilliams}} parameterization.
\newblock \emph{Ocean Modelling}, 136:\penalty0 85--106, Apr. 2019.
\newblock ISSN 1463-5003.
\newblock \doi{10.1016/j.ocemod.2019.02.006}.

\bibitem[Bachman et~al.(2017{\natexlab{a}})Bachman, {Fox-Kemper}, and
  Pearson]{Bachmanetal2017}
S.~D. Bachman, B.~{Fox-Kemper}, and B.~Pearson.
\newblock A scale-aware subgrid model for quasi-geostrophic turbulence.
\newblock \emph{Journal of Geophysical Research: Oceans}, 122\penalty0
  (2):\penalty0 1529--1554, 2017{\natexlab{a}}.
\newblock ISSN 2169-9291.
\newblock \doi{10.1002/2016JC012265}.

\bibitem[Bachman et~al.(2017{\natexlab{b}})Bachman, Marshall, Maddison, and
  Mak]{bachmanEvaluationScalarEddy2017}
S.~D. Bachman, D.~P. Marshall, J.~R. Maddison, and J.~Mak.
\newblock Evaluation of a scalar eddy transport coefficient based on geometric
  constraints.
\newblock \emph{Ocean Modelling}, 109:\penalty0 44--54, Jan.
  2017{\natexlab{b}}.
\newblock ISSN 1463-5003.
\newblock \doi{10.1016/j.ocemod.2016.12.004}.

\bibitem[Balwada et~al.(2026)Balwada, Perezhogin, Adcroft, and
  Zanna]{balwadaDesignImplementationDataDriven2026}
D.~Balwada, P.~Perezhogin, A.~Adcroft, and L.~Zanna.
\newblock Design and {{Implementation}} of a {{Data-Driven Parameterization}}
  for {{Mesoscale Eddy-Induced Transport}}: {{Idealized Two-Layer
  Configurations}}.
\newblock \emph{Journal of Advances in Modeling Earth Systems}, 18\penalty0
  (9):\penalty0 e2025MS005245, 2026.
\newblock ISSN 1942-2466.
\newblock \doi{10.1029/2025MS005245}.

\bibitem[Beucler et~al.(2021)Beucler, Pritchard, Rasp, Ott, Baldi, and
  Gentine]{beuclerEnforcingAnalyticConstraints2021}
T.~Beucler, M.~Pritchard, S.~Rasp, J.~Ott, P.~Baldi, and P.~Gentine.
\newblock Enforcing {{Analytic Constraints}} in {{Neural Networks Emulating
  Physical Systems}}.
\newblock \emph{Physical Review Letters}, 126\penalty0 (9):\penalty0 098302,
  Mar. 2021.
\newblock \doi{10.1103/PhysRevLett.126.098302}.

\bibitem[Bodner et~al.(2025)Bodner, Balwada, and
  Zanna]{bodnerDataDrivenApproachParameterizing2025a}
A.~Bodner, D.~Balwada, and L.~Zanna.
\newblock A {{Data-Driven Approach}} for {{Parameterizing Ocean Submesoscale
  Buoyancy Fluxes}}.
\newblock \emph{Journal of Advances in Modeling Earth Systems}, 17\penalty0
  (11):\penalty0 e2025MS004991, 2025.
\newblock ISSN 1942-2466.
\newblock \doi{10.1029/2025MS004991}.

\bibitem[Bolton and Zanna(2019)]{boltonApplicationsDeepLearning2019}
T.~Bolton and L.~Zanna.
\newblock Applications of {{Deep Learning}} to {{Ocean Data Inference}} and
  {{Subgrid Parameterization}}.
\newblock \emph{Journal of Advances in Modeling Earth Systems}, 11\penalty0
  (1):\penalty0 376--399, 2019.
\newblock ISSN 1942-2466.
\newblock \doi{10.1029/2018MS001472}.

\bibitem[Buzzicotti et~al.(2023)Buzzicotti, Storer, Khatri, Griffies, and
  Aluie]{buzzicottiSpatioTemporalCoarseGrainingDecomposition2023}
M.~Buzzicotti, B.~A. Storer, H.~Khatri, S.~M. Griffies, and H.~Aluie.
\newblock Spatio-{{Temporal Coarse-Graining Decomposition}} of the {{Global
  Ocean Geostrophic Kinetic Energy}}.
\newblock \emph{Journal of Advances in Modeling Earth Systems}, 15\penalty0
  (6):\penalty0 e2023MS003693, 2023.
\newblock ISSN 1942-2466.
\newblock \doi{10.1029/2023MS003693}.

\bibitem[Chattopadhyay and
  Hassanzadeh(2023)]{chattopadhyayLongtermInstabilitiesDeep2023}
A.~Chattopadhyay and P.~Hassanzadeh.
\newblock Long-term instabilities of deep learning-based digital twins of the
  climate system: {{The}} cause and a solution, Apr. 2023.

\bibitem[Chelton et~al.(2011)Chelton, Schlax, and Samelson]{Chelton2011}
D.~B. Chelton, M.~G. Schlax, and R.~M. Samelson.
\newblock Global observations of nonlinear mesoscale eddies.
\newblock \emph{Progress in Oceanography}, 91\penalty0 (2):\penalty0 167--216,
  Oct. 2011.
\newblock ISSN 0079-6611.
\newblock \doi{10.1016/j.pocean.2011.01.002}.

\bibitem[Chen et~al.(2019)Chen, Rubanova, Bettencourt, and
  Duvenaud]{chenNeuralOrdinaryDifferential2019}
R.~T.~Q. Chen, Y.~Rubanova, J.~Bettencourt, and D.~Duvenaud.
\newblock Neural {{Ordinary Differential Equations}}, Dec. 2019.

\bibitem[Cl{\'e}ment(2021)]{clementCommunityGlobal12021}
B.~Cl{\'e}ment.
\newblock Toward a community global 1/36{$^\circ$} {{ORCA36}} configuration
  based on {{NEMO}} 4.
\newblock Oct. 2021.
\newblock \doi{10.5281/zenodo.5571815}.

\bibitem[Darman et~al.(2026)Darman, Hassanzadeh, Zanna, and
  Chattopadhyay]{darmanFourierAnalysisPhysics2026}
M.~Darman, P.~Hassanzadeh, L.~Zanna, and A.~Chattopadhyay.
\newblock Fourier analysis of the physics of transfer learning for data-driven
  subgrid-scale models of ocean turbulence.
\newblock \emph{Machine Learning: Earth}, Apr. 2026.
\newblock \doi{10.1088/3049-4753/ae510d}.

\bibitem[{de La Lama} et~al.(2016){de La Lama}, LaCasce, and
  Fuhr]{delalama2016}
M.~S. {de La Lama}, J.~H. LaCasce, and H.~K. Fuhr.
\newblock The vertical structure of ocean eddies.
\newblock \emph{Dynamics and Statistics of the Climate System}, 1\penalty0
  (1):\penalty0 dzw001, Jan. 2016.
\newblock ISSN 2059-6987.
\newblock \doi{10.1093/climsys/dzw001}.

\bibitem[Dunne et~al.(2025)Dunne, Hewitt, Arblaster, Bonou, Boucher, Cavazos,
  Dingley, Durack, Hassler, Juckes, Miyakawa, Mizielinski, Naik, Nicholls,
  O'Rourke, Pincus, Sanderson, Simpson, and
  Taylor]{dunneEvolvingCoupledModel2025}
J.~P. Dunne, H.~T. Hewitt, J.~M. Arblaster, F.~Bonou, O.~Boucher, T.~Cavazos,
  B.~Dingley, P.~J. Durack, B.~Hassler, M.~Juckes, T.~Miyakawa, M.~Mizielinski,
  V.~Naik, Z.~Nicholls, E.~O'Rourke, R.~Pincus, B.~M. Sanderson, I.~R. Simpson,
  and K.~E. Taylor.
\newblock An evolving {{Coupled Model Intercomparison Project}} phase 7
  ({{CMIP7}}) and {{Fast Track}} in support of future climate assessment.
\newblock \emph{Geoscientific Model Development}, 18\penalty0 (19):\penalty0
  6671--6700, Oct. 2025.
\newblock ISSN 1991-959X.
\newblock \doi{10.5194/gmd-18-6671-2025}.

\bibitem[Durand et~al.(2024)Durand, Finn, Farchi, Bocquet, Boutin, and
  {\'O}lason]{durandDatadrivenSurrogateModeling2024}
C.~Durand, T.~S. Finn, A.~Farchi, M.~Bocquet, G.~Boutin, and E.~{\'O}lason.
\newblock Data-driven surrogate modeling of high-resolution sea-ice thickness
  in the {{Arctic}}.
\newblock \emph{The Cryosphere}, 18\penalty0 (4):\penalty0 1791--1815, Apr.
  2024.
\newblock ISSN 1994-0416.
\newblock \doi{10.5194/tc-18-1791-2024}.

\bibitem[Ferreira et~al.(2005)Ferreira, Marshall, and Heimbach]{Ferreira2005}
D.~Ferreira, J.~Marshall, and P.~Heimbach.
\newblock Estimating {{Eddy Stresses}} by {{Fitting Dynamics}} to
  {{Observations Using}} a {{Residual-Mean Ocean Circulation Model}} and {{Its
  Adjoint}}.
\newblock \emph{Journal of Physical Oceanography}, 35\penalty0 (10):\penalty0
  1891--1910, Oct. 2005.
\newblock ISSN 0022-3670, 1520-0485.
\newblock \doi{10.1175/JPO2785.1}.

\bibitem[{Fox-Kemper} et~al.(2019){Fox-Kemper}, Adcroft, B{\"o}ning,
  Chassignet, Curchitser, Danabasoglu, Eden, England, Gerdes, Greatbatch,
  Griffies, Hallberg, Hanert, Heimbach, Hewitt, Hill, Komuro, Legg, Le~Sommer,
  Masina, Marsland, Penny, Qiao, Ringler, Treguier, Tsujino, Uotila, and
  Yeager]{fox-kemperChallengesProspectsOcean2019}
B.~{Fox-Kemper}, A.~Adcroft, C.~W. B{\"o}ning, E.~P. Chassignet, E.~Curchitser,
  G.~Danabasoglu, C.~Eden, M.~H. England, R.~Gerdes, R.~J. Greatbatch, S.~M.
  Griffies, R.~W. Hallberg, E.~Hanert, P.~Heimbach, H.~T. Hewitt, C.~N. Hill,
  Y.~Komuro, S.~Legg, J.~Le~Sommer, S.~Masina, S.~J. Marsland, S.~G. Penny,
  F.~Qiao, T.~D. Ringler, A.~M. Treguier, H.~Tsujino, P.~Uotila, and S.~G.
  Yeager.
\newblock Challenges and {{Prospects}} in {{Ocean Circulation Models}}.
\newblock \emph{Frontiers in Marine Science}, 6, 2019.
\newblock ISSN 2296-7745.

\bibitem[Frezat et~al.(2022)Frezat, Le~Sommer, Fablet, Balarac, and
  Lguensat]{frezatPosterioriLearningQuasiGeostrophic2022}
H.~Frezat, J.~Le~Sommer, R.~Fablet, G.~Balarac, and R.~Lguensat.
\newblock A {{Posteriori Learning}} for {{Quasi-Geostrophic Turbulence
  Parametrization}}.
\newblock \emph{Journal of Advances in Modeling Earth Systems}, 14\penalty0
  (11):\penalty0 e2022MS003124, 2022.
\newblock ISSN 1942-2466.
\newblock \doi{10.1029/2022MS003124}.

\bibitem[Frigola et~al.(2025)Frigola, {Martin-Martinez}, {Moreno-Chamarro},
  Sams{\'o}, {Loosvelt-Tomas}, Bretonni{\`e}re, Kuznetsova, Lin, and
  Ortega]{frigolaNorthAtlanticMean2025a}
A.~Frigola, E.~{Martin-Martinez}, E.~{Moreno-Chamarro}, M.~Sams{\'o},
  S.~{Loosvelt-Tomas}, P.-A. Bretonni{\`e}re, D.~Kuznetsova, X.~Lin, and
  P.~Ortega.
\newblock The {{North Atlantic}} mean state in mesoscale eddy-resolving coupled
  models: A multimodel study.
\newblock \emph{Ocean Science}, 21\penalty0 (6):\penalty0 3507--3540, Dec.
  2025.
\newblock ISSN 1812-0784.
\newblock \doi{10.5194/os-21-3507-2025}.

\bibitem[Furtado et~al.(2026)Furtado, Molina, Arcodia, Anderson, Beucler,
  Callahan, Ciasto, Gensini, L'Heureux, Pegion, {P{\'e}rez-Carrasquilla},
  Sonnewald, Takahashi, Xiang, and
  Zimmerman]{furtadoSettingStandardRecommended2026}
J.~C. Furtado, M.~J. Molina, M.~C. Arcodia, W.~Anderson, T.~Beucler, J.~A.
  Callahan, L.~M. Ciasto, V.~A. Gensini, M.~L'Heureux, K.~Pegion, J.~S.
  {P{\'e}rez-Carrasquilla}, M.~Sonnewald, K.~Takahashi, B.~Xiang, and B.~G.
  Zimmerman.
\newblock Setting the {{Standard}}: {{Recommended Practices}} for {{Data
  Preprocessing}} in {{Data-Driven Climate Prediction}}.
\newblock \emph{Bulletin of the American Meteorological Society}, -1\penalty0
  (aop), Apr. 2026.
\newblock ISSN 0003-0007, 1520-0477.
\newblock \doi{10.1175/BAMS-D-24-0292.1}.

\bibitem[Gent and McWilliams(1990)]{GM1990}
P.~R. Gent and J.~C. McWilliams.
\newblock Isopycnal {{Mixing}} in {{Ocean Circulation Models}}.
\newblock \emph{Journal of Physical Oceanography}, 20\penalty0 (1):\penalty0
  150--155, Jan. 1990.
\newblock ISSN 0022-3670, 1520-0485.
\newblock \doi{10.1175/1520-0485(1990)020<0150:IMIOCM>2.0.CO;2}.

\bibitem[Gentine et~al.(2026)Gentine, Balwada, Pa{\c c}al, Hawkins, Adcroft,
  Fan, Fang, Lahlou, Lamb, Lawrence, Sommer, Mouallem, Nathaniel,
  {Watson-Parris}, and Eyring]{gentineLegoESMModularDifferentiable2026}
P.~Gentine, D.~Balwada, A.~Pa{\c c}al, L.~Hawkins, A.~Adcroft, H.~Fan, J.~Fang,
  A.~Lahlou, K.~D. Lamb, D.~M. Lawrence, J.~L. Sommer, J.~Mouallem,
  J.~Nathaniel, D.~{Watson-Parris}, and V.~Eyring.
\newblock {{legoESM}}: A modular, differentiable, multiscale, {{AI}}-ready
  {{Earth}} system model built with {{AI}} agents, Oct. 2026.

\bibitem[Ghosh et~al.(2026)Ghosh, Shukla, Singh, and
  Chanda]{ghoshPhysicsembeddedTransformerCNNArchitecture2026}
S.~Ghosh, V.~K. Shukla, A.~Singh, and J.~Chanda.
\newblock A physics-embedded {{Transformer-CNN}} architecture for data-driven
  turbulence prediction and surrogate modeling of high-fidelity fluid dynamics.
\newblock \emph{European Journal of Mechanics - B/Fluids}, 115:\penalty0
  204372, Jan. 2026.
\newblock ISSN 0997-7546.
\newblock \doi{10.1016/j.euromechflu.2025.204372}.

\bibitem[Guan et~al.(2026)Guan, Amoudruz, Litvinov, Jakhar, Mojgani,
  Koumoutsakos, and Hassanzadeh]{guanPredictionExtremeEvents2026}
Y.~Guan, L.~Amoudruz, S.~Litvinov, K.~Jakhar, R.~Mojgani, P.~Koumoutsakos, and
  P.~Hassanzadeh.
\newblock Prediction of {{Extreme Events}} in {{Multiscale Simulations}} of
  {{Geophysical Turbulence}} using {{Reinforcement Learning}}, Feb. 2026.

\bibitem[Hallberg(2013)]{Hallberg2013}
R.~Hallberg.
\newblock Using a resolution function to regulate parameterizations of oceanic
  mesoscale eddy effects.
\newblock \emph{Ocean Modelling}, 72:\penalty0 92--103, Dec. 2013.
\newblock ISSN 1463-5003.
\newblock \doi{10.1016/j.ocemod.2013.08.007}.

\bibitem[Heuer et~al.(2025)Heuer, Beucler, Schwabe, Savre, Schlund, and
  Eyring]{heuerTrainingDataConfidenceGuided2025}
H.~Heuer, T.~Beucler, M.~Schwabe, J.~Savre, M.~Schlund, and V.~Eyring.
\newblock Beyond the {{Training Data}}: {{Confidence-Guided Mixing}} of
  {{Parameterizations}} in a {{Hybrid AI-Climate Model}}, 2025.

\bibitem[Hewitt et~al.(2022)Hewitt, {Fox-Kemper}, Pearson, Roberts, and
  Klocke]{Hewitt2022}
H.~Hewitt, B.~{Fox-Kemper}, B.~Pearson, M.~Roberts, and D.~Klocke.
\newblock The small scales of the ocean may hold the key to surprises.
\newblock \emph{Nature Climate Change}, 12\penalty0 (6):\penalty0 496--499,
  June 2022.
\newblock ISSN 1758-6798.
\newblock \doi{10.1038/s41558-022-01386-6}.

\bibitem[Hu et~al.(2025)Hu, Subramaniam, Kuang, Lin, Yu, Hannah, Brenowitz,
  Romero, and Pritchard]{huStableMachineLearningParameterization2025}
Z.~Hu, A.~Subramaniam, Z.~Kuang, J.~Lin, S.~Yu, W.~M. Hannah, N.~D. Brenowitz,
  J.~Romero, and M.~S. Pritchard.
\newblock Stable {{Machine-Learning Parameterization}} of {{Subgrid Processes}}
  in a {{Comprehensive Atmospheric Model Learned From Embedded
  Convection-Permitting Simulations}}.
\newblock \emph{Journal of Advances in Modeling Earth Systems}, 17\penalty0
  (7):\penalty0 e2024MS004618, 2025.
\newblock ISSN 1942-2466.
\newblock \doi{10.1029/2024MS004618}.

\bibitem[Jackson et~al.(2023)Jackson, Hewitt, Bruciaferri, Calvert, Graham,
  Guiavarc'h, Menary, New, Roberts, and
  Storkey]{jacksonChallengesSimulatingAMOC2023}
L.~C. Jackson, H.~T. Hewitt, D.~Bruciaferri, D.~Calvert, T.~Graham,
  C.~Guiavarc'h, M.~B. Menary, A.~L. New, M.~Roberts, and D.~Storkey.
\newblock Challenges simulating the {{AMOC}} in climate models.
\newblock \emph{Philosophical Transactions of the Royal Society A:
  Mathematical, Physical and Engineering Sciences}, 381\penalty0
  (2262):\penalty0 20220187, Oct. 2023.
\newblock ISSN 1364-503X.
\newblock \doi{10.1098/rsta.2022.0187}.

\bibitem[Jakhar et~al.(2024)Jakhar, Guan, Mojgani, Chattopadhyay, and
  Hassanzadeh]{jakharLearningClosedFormEquations2024}
K.~Jakhar, Y.~Guan, R.~Mojgani, A.~Chattopadhyay, and P.~Hassanzadeh.
\newblock Learning {{Closed-Form Equations}} for {{Subgrid-Scale Closures From
  High-Fidelity Data}}: {{Promises}} and {{Challenges}}.
\newblock \emph{Journal of Advances in Modeling Earth Systems}, 16\penalty0
  (7):\penalty0 e2023MS003874, 2024.
\newblock ISSN 1942-2466.
\newblock \doi{10.1029/2023MS003874}.

\bibitem[Jakhar et~al.(2026)Jakhar, Guan, and
  Hassanzadeh]{jakharAnalyticalAIDiscoveredStable2026}
K.~Jakhar, Y.~Guan, and P.~Hassanzadeh.
\newblock Analytical and {{AI-Discovered Stable}}, {{Accurate}}, and
  {{Generalizable Subgrid-Scale Closure}} for {{Geophysical Turbulence}}.
\newblock \emph{Physical Review Letters}, 136\penalty0 (6):\penalty0 064201,
  Feb. 2026.
\newblock ISSN 0031-9007, 1079-7114.
\newblock \doi{10.1103/v28b-5qmp}.

\bibitem[Jansen et~al.(2019)Jansen, Adcroft, Khani, and Kong]{Jansen2019}
M.~F. Jansen, A.~Adcroft, S.~Khani, and H.~Kong.
\newblock Toward an {{Energetically Consistent}}, {{Resolution Aware
  Parameterization}} of {{Ocean Mesoscale Eddies}}.
\newblock \emph{Journal of Advances in Modeling Earth Systems}, 11\penalty0
  (8):\penalty0 2844--2860, 2019.
\newblock ISSN 1942-2466.
\newblock \doi{10.1029/2019MS001750}.

\bibitem[Kamm et~al.(2025)Kamm, Deshayes, and
  Madec]{kammDINODiabaticModel2025a}
D.~Kamm, J.~Deshayes, and G.~Madec.
\newblock {{DINO}}: A diabatic model of pole-to-pole ocean dynamics to assess
  subgrid parameterizations across horizontal scales.
\newblock \emph{Geoscientific Model Development}, 18\penalty0 (21):\penalty0
  8091--8107, Oct. 2025.
\newblock ISSN 1991-959X.
\newblock \doi{10.5194/gmd-18-8091-2025}.

\bibitem[Loose et~al.(2022)Loose, Abernathey, Grooms, Busecke, Guillaumin,
  Yankovsky, Marques, Steinberg, Ross, Khatri, Bachman, Zanna, and
  Martin]{looseGCMFiltersPythonPackage2022}
N.~Loose, R.~Abernathey, I.~Grooms, J.~Busecke, A.~Guillaumin, E.~Yankovsky,
  G.~Marques, J.~Steinberg, A.~S. Ross, H.~Khatri, S.~Bachman, L.~Zanna, and
  P.~Martin.
\newblock {{GCM-Filters}}: {{A Python Package}} for {{Diffusion}}-based
  {{Spatial Filtering}} of {{Gridded Data}}.
\newblock \emph{Journal of Open Source Software}, 7\penalty0 (70):\penalty0
  3947, Feb. 2022.
\newblock ISSN 2475-9066.
\newblock \doi{10.21105/joss.03947}.

\bibitem[Madec et~al.(2023)Madec, Bell, Blaker, Bricaud, Bruciaferri,
  Castrillo, Calvert, Chanut, Clementi, Coward, Epicoco, {\'E}th{\'e},
  Ganderton, Harle, Hutchinson, Iovino, Lea, Lovato, Martin, Martin, Mele,
  Martins, Masson, Mathiot, Mele, Mocavero, M{\"u}ller, Nurser, Paronuzzi,
  Peltier, Person, Rousset, Rynders, Samson, T{\'e}chen{\'e}, Vancoppenolle,
  and Wilson]{Madec2023}
G.~Madec, M.~Bell, A.~Blaker, C.~Bricaud, D.~Bruciaferri, M.~Castrillo,
  D.~Calvert, J.~Chanut, E.~Clementi, A.~Coward, I.~Epicoco, C.~{\'E}th{\'e},
  J.~Ganderton, J.~Harle, K.~Hutchinson, D.~Iovino, D.~Lea, T.~Lovato,
  M.~Martin, N.~Martin, F.~Mele, D.~Martins, S.~Masson, P.~Mathiot, F.~Mele,
  S.~Mocavero, S.~M{\"u}ller, A.~J.~G. Nurser, S.~Paronuzzi, M.~Peltier,
  R.~Person, C.~Rousset, S.~Rynders, G.~Samson, S.~T{\'e}chen{\'e},
  M.~Vancoppenolle, and C.~Wilson.
\newblock {{NEMO Ocean Engine Reference Manual}}.
\newblock July 2023.
\newblock \doi{10.5281/zenodo.8167700}.

\bibitem[Mak et~al.(2018)Mak, Maddison, Marshall, and Munday]{Mak2018}
J.~Mak, J.~R. Maddison, D.~P. Marshall, and D.~R. Munday.
\newblock Implementation of a {{Geometrically Informed}} and {{Energetically
  Constrained Mesoscale Eddy Parameterization}} in an {{Ocean Circulation
  Model}}.
\newblock \emph{Journal of Physical Oceanography}, 48\penalty0 (10):\penalty0
  2363--2382, Oct. 2018.
\newblock ISSN 0022-3670, 1520-0485.
\newblock \doi{10.1175/JPO-D-18-0017.1}.

\bibitem[Mak et~al.(2023)Mak, Maddison, Marshall, Ruan, Wang, and
  Yeow]{makScaleAwarenessEddyEnergy2023}
J.~Mak, J.~R. Maddison, D.~Marshall, X.~Ruan, Y.~Wang, and L.~Yeow.
\newblock Scale-{{Awareness}} in an {{Eddy Energy Constrained Mesoscale Eddy
  Parameterization}}.
\newblock \emph{Journal of Advances in Modeling Earth Systems}, 15\penalty0
  (12):\penalty0 e2023MS003886, 2023.
\newblock ISSN 1942-2466.
\newblock \doi{10.1029/2023MS003886}.

\bibitem[Marques et~al.(2022)Marques, Loose, Yankovsky, Steinberg, Chang,
  Bhamidipati, Adcroft, {Fox-Kemper}, Griffies, Hallberg, Jansen, Khatri, and
  Zanna]{marquesNeverWorld2IdealizedModel2022}
G.~M. Marques, N.~Loose, E.~Yankovsky, J.~M. Steinberg, C.-Y. Chang,
  N.~Bhamidipati, A.~Adcroft, B.~{Fox-Kemper}, S.~M. Griffies, R.~W. Hallberg,
  M.~F. Jansen, H.~Khatri, and L.~Zanna.
\newblock {{NeverWorld2}}: An idealized model hierarchy to investigate ocean
  mesoscale eddies across resolutions.
\newblock \emph{Geoscientific Model Development}, 15\penalty0 (17):\penalty0
  6567--6579, Sept. 2022.
\newblock ISSN 1991-959X.
\newblock \doi{10.5194/gmd-15-6567-2022}.

\bibitem[Marshall et~al.(2012)Marshall, Maddison, and Berloff]{Marshall2012}
D.~Marshall, J.~R. Maddison, and P.~S. Berloff.
\newblock A {{Framework}} for {{Parameterizing Eddy Potential Vorticity
  Fluxes}}.
\newblock \emph{Journal of Physical Oceanography}, 42\penalty0 (4):\penalty0
  539--557, Apr. 2012.
\newblock ISSN 0022-3670, 1520-0485.
\newblock \doi{10.1175/JPO-D-11-048.1}.

\bibitem[Maulik et~al.(2019)Maulik, San, Rasheed, and
  Vedula]{maulikSubgridModellingTwodimensional2019}
R.~Maulik, O.~San, A.~Rasheed, and P.~Vedula.
\newblock Subgrid modelling for two-dimensional turbulence using neural
  networks.
\newblock \emph{Journal of Fluid Mechanics}, 858:\penalty0 122--144, Jan. 2019.
\newblock ISSN 0022-1120, 1469-7645.
\newblock \doi{10.1017/jfm.2018.770}.

\bibitem[Nath et~al.(2026)Nath, Schemm, Moss, Haynes, Shuckburgh, and
  Webb]{nathReplacingTunableParameters2026}
P.~Nath, S.~Schemm, H.~Moss, P.~Haynes, E.~Shuckburgh, and M.~J. Webb.
\newblock Replacing {{Tunable Parameters}} in {{Weather}} and {{Climate Models
  With State-Dependent Functions Using Reinforcement Learning}}.
\newblock \emph{Journal of Advances in Modeling Earth Systems}, 18\penalty0
  (8):\penalty0 e2026MS005745, 2026.
\newblock ISSN 1942-2466.
\newblock \doi{10.1029/2026MS005745}.

\bibitem[Nowak et~al.(2025)Nowak, Danilov, M{\"u}ller, and
  Liu]{nowakImplementationImplicitFilters2025}
K.~Nowak, S.~Danilov, V.~M{\"u}ller, and C.~Liu.
\newblock Implementation of implicit filters for spatial spectra extraction.
\newblock \emph{Geoscientific Model Development}, 18\penalty0 (18):\penalty0
  6541--6551, Sept. 2025.
\newblock ISSN 1991-959X.
\newblock \doi{10.5194/gmd-18-6541-2025}.

\bibitem[O'Gorman and Dwyer(2018)]{ogormanUsingMachineLearning2018}
P.~A. O'Gorman and J.~G. Dwyer.
\newblock Using {{Machine Learning}} to {{Parameterize Moist Convection}}:
  {{Potential}} for {{Modeling}} of {{Climate}}, {{Climate Change}}, and
  {{Extreme Events}}.
\newblock \emph{Journal of Advances in Modeling Earth Systems}, 10\penalty0
  (10):\penalty0 2548--2563, 2018.
\newblock ISSN 1942-2466.
\newblock \doi{10.1029/2018MS001351}.

\bibitem[Ouala et~al.(2024)Ouala, Chapron, Collard, Gaultier, and
  Fablet]{oualaOnlineCalibrationDeep2024}
S.~Ouala, B.~Chapron, F.~Collard, L.~Gaultier, and R.~Fablet.
\newblock Online calibration of deep learning sub-models for hybrid numerical
  modeling systems.
\newblock \emph{Communications Physics}, 7\penalty0 (1):\penalty0 1--15, Dec.
  2024.
\newblock ISSN 2399-3650.
\newblock \doi{10.1038/s42005-024-01880-7}.

\bibitem[Partee et~al.(2022)Partee, Ellis, Rigazzi, Shao, Bachman, Marques, and
  Robbins]{parteeUsingMachineLearning2022}
S.~Partee, M.~Ellis, A.~Rigazzi, A.~E. Shao, S.~Bachman, G.~Marques, and
  B.~Robbins.
\newblock Using {{Machine Learning}} at scale in numerical simulations with
  {{SmartSim}}: {{An}} application to ocean climate modeling.
\newblock \emph{Journal of Computational Science}, 62:\penalty0 101707, July
  2022.
\newblock ISSN 1877-7503.
\newblock \doi{10.1016/j.jocs.2022.101707}.

\bibitem[Perezhogin et~al.(2023)Perezhogin, Zanna, and
  {Fernandez-Granda}]{perezhoginGenerativeDataDrivenApproaches2023}
P.~Perezhogin, L.~Zanna, and C.~{Fernandez-Granda}.
\newblock Generative {{Data-Driven Approaches}} for {{Stochastic Subgrid
  Parameterizations}} in an {{Idealized Ocean Model}}.
\newblock \emph{Journal of Advances in Modeling Earth Systems}, 15\penalty0
  (10):\penalty0 e2023MS003681, 2023.
\newblock ISSN 1942-2466.
\newblock \doi{10.1029/2023MS003681}.

\bibitem[Perezhogin et~al.(2024)Perezhogin, Zhang, Adcroft, {Fernandez-Granda},
  and Zanna]{perezhoginStableImplementationDataDriven2024}
P.~Perezhogin, C.~Zhang, A.~Adcroft, C.~{Fernandez-Granda}, and L.~Zanna.
\newblock A {{Stable Implementation}} of a {{Data-Driven Scale-Aware Mesoscale
  Parameterization}}.
\newblock \emph{Journal of Advances in Modeling Earth Systems}, 16\penalty0
  (10):\penalty0 e2023MS004104, 2024.
\newblock ISSN 1942-2466.
\newblock \doi{10.1029/2023MS004104}.

\bibitem[Perezhogin et~al.(2025)Perezhogin, Adcroft, and
  Zanna]{perezhoginGeneralizableNeuralNetworkParameterization2025}
P.~Perezhogin, A.~Adcroft, and L.~Zanna.
\newblock Generalizable {{Neural-Network Parameterization}} of {{Mesoscale
  Eddies}} in {{Idealized}} and {{Global Ocean Models}}.
\newblock \emph{Geophysical Research Letters}, 52\penalty0 (19):\penalty0
  e2025GL117046, 2025.
\newblock ISSN 1944-8007.
\newblock \doi{10.1029/2025GL117046}.

\bibitem[Rasp et~al.(2018)Rasp, Pritchard, and
  Gentine]{raspDeepLearningRepresent2018}
S.~Rasp, M.~S. Pritchard, and P.~Gentine.
\newblock Deep learning to represent subgrid processes in climate models.
\newblock \emph{Proceedings of the National Academy of Sciences}, 115\penalty0
  (39):\penalty0 9684--9689, Sept. 2018.
\newblock \doi{10.1073/pnas.1810286115}.

\bibitem[Raut et~al.(2026)Raut, Gupta, Jackson, Wang, Theisen, Tian, Park,
  O'Brien, Grover, Arcomano, Hickmon, Collis, Ferrier, and
  Dematties]{rautPerspectiveSelfSupervisedTransformers2026}
B.~A. Raut, S.~Gupta, R.~C. Jackson, D.~Wang, A.~Theisen, J.~Tian, S.~Park,
  J.~R. O'Brien, M.~A. Grover, T.~Arcomano, N.~L. Hickmon, S.~M. Collis, N.~J.
  Ferrier, and D.~J. Dematties.
\newblock Perspective on {{Self-Supervised Transformers}} for {{Scalable Data
  Analysis}}, {{Retrievals}}, and {{Discovery}} in {{Geosciences}}.
\newblock \emph{Artificial Intelligence for the Earth Systems}, 5\penalty0 (4),
  Aug. 2026.
\newblock ISSN 2769-7525.
\newblock \doi{10.1175/AIES-D-25-0069.1}.

\bibitem[Ronneberger et~al.(2015)Ronneberger, Fischer, and
  Brox]{ronnebergerUNetConvolutionalNetworks2015}
O.~Ronneberger, P.~Fischer, and T.~Brox.
\newblock U-{{Net}}: {{Convolutional Networks}} for {{Biomedical Image
  Segmentation}}, May 2015.

\bibitem[Ross et~al.(2023)Ross, Li, Perezhogin, {Fernandez-Granda}, and
  Zanna]{rossBenchmarkingMachineLearning2023}
A.~Ross, Z.~Li, P.~Perezhogin, C.~{Fernandez-Granda}, and L.~Zanna.
\newblock Benchmarking of {{Machine Learning Ocean Subgrid Parameterizations}}
  in an {{Idealized Model}}.
\newblock \emph{Journal of Advances in Modeling Earth Systems}, 15\penalty0
  (1):\penalty0 e2022MS003258, 2023.
\newblock ISSN 1942-2466.
\newblock \doi{10.1029/2022MS003258}.

\bibitem[{Saynisch-Wagner} and
  Sari(2026)]{saynisch-wagnerConditionalUpdatesNeural2026}
J.~{Saynisch-Wagner} and S.~R. Sari.
\newblock Conditional updates of neural network weights for increased out of
  training performance.
\newblock \emph{Nonlinear Processes in Geophysics}, 33\penalty0 (4):\penalty0
  503--520, Oct. 2026.
\newblock ISSN 1023-5809.
\newblock \doi{10.5194/npg-33-503-2026}.

\bibitem[Segura et~al.(2025)Segura, {Pedruzo-Bagazgoitia}, Weiss, M{\"u}ller,
  Rackow, Lee, {Dolores-Tesillos}, Benedict, Aengenheyster, Aguridan, Arduini,
  Baker, Bao, Bastin, Baulenas, Becker, Beyer, Bockelmann, Br{\"u}ggemann,
  Brunner, Cheedela, Das, Denissen, Dragaud, Dziekan, Ekblom, Engels, Esch,
  Forbes, Frauen, Freischem, {Garc{\'i}a-Maroto}, Geier, Gierz,
  {Gonz{\'a}lez-Cervera}, Grayson, Griffith, Gutjahr, Haak, Hadade, Haslehner,
  {ul Hasson}, Hegewald, Kluft, Koldunov, Koldunov, K{\"o}lling, Koseki,
  Kosukhin, Kousal, Kuma, Kumar, Li, Maury, Meindl, Milinski, Mogensen,
  Niraula, Nowak, Praturi, Proske, Putrasahan, Redler, Santuy, S{\'a}rm{\'a}ny,
  Schnur, Scholz, Sidorenko, Sp{\"a}t, S{\"u}tzl, Takasuka, Tompkins, Uribe,
  Valentini, Veerman, Voigt, Warnau, Wachsmann, Wac{\l}awczyk, Wedi, Wieners,
  Wille, Winkler, Wu, Ziemen, Zimmermann, Bender, Bojovic, Bony, Bordoni,
  Brehmer, Dengler, Dutra, Faye, Fischer, {van Heerwaarden}, Hohenegger,
  J{\"a}rvinen, Jochum, Jung, Jungclaus, Keenlyside, Klocke, Konow, Klose,
  Malinowski, Martius, Mauritsen, Mellado, Mieslinger, Mohino, Paw{\l}owska,
  {Peters-von Gehlen}, Sarr{\'e}, Sobhani, Stier, Tuppi, Vidale, Sandu, and
  Stevens]{seguraNextGEMSEnteringEra2025a}
H.~Segura, X.~{Pedruzo-Bagazgoitia}, P.~Weiss, S.~K. M{\"u}ller, T.~Rackow,
  J.~Lee, E.~{Dolores-Tesillos}, I.~Benedict, M.~Aengenheyster, R.~Aguridan,
  G.~Arduini, A.~J. Baker, J.~Bao, S.~Bastin, E.~Baulenas, T.~Becker, S.~Beyer,
  H.~Bockelmann, N.~Br{\"u}ggemann, L.~Brunner, S.~K. Cheedela, S.~Das,
  J.~Denissen, I.~Dragaud, P.~Dziekan, M.~Ekblom, J.~F. Engels, M.~Esch,
  R.~Forbes, C.~Frauen, L.~Freischem, D.~{Garc{\'i}a-Maroto}, P.~Geier,
  P.~Gierz, {\'A}.~{Gonz{\'a}lez-Cervera}, K.~Grayson, M.~Griffith, O.~Gutjahr,
  H.~Haak, I.~Hadade, K.~Haslehner, S.~{ul Hasson}, J.~Hegewald, L.~Kluft,
  A.~Koldunov, N.~Koldunov, T.~K{\"o}lling, S.~Koseki, S.~Kosukhin, J.~Kousal,
  P.~Kuma, A.~U. Kumar, R.~Li, N.~Maury, M.~Meindl, S.~Milinski, K.~Mogensen,
  B.~Niraula, J.~Nowak, D.~S. Praturi, U.~Proske, D.~Putrasahan, R.~Redler,
  D.~Santuy, D.~S{\'a}rm{\'a}ny, R.~Schnur, P.~Scholz, D.~Sidorenko,
  D.~Sp{\"a}t, B.~S{\"u}tzl, D.~Takasuka, A.~Tompkins, A.~Uribe, M.~Valentini,
  M.~Veerman, A.~Voigt, S.~Warnau, F.~Wachsmann, M.~Wac{\l}awczyk, N.~Wedi,
  K.-H. Wieners, J.~Wille, M.~Winkler, Y.~Wu, F.~Ziemen, J.~Zimmermann,
  F.~A.-M. Bender, D.~Bojovic, S.~Bony, S.~Bordoni, P.~Brehmer, M.~Dengler,
  E.~Dutra, S.~Faye, E.~Fischer, C.~{van Heerwaarden}, C.~Hohenegger,
  H.~J{\"a}rvinen, M.~Jochum, T.~Jung, J.~H. Jungclaus, N.~S. Keenlyside,
  D.~Klocke, H.~Konow, M.~Klose, S.~Malinowski, O.~Martius, T.~Mauritsen, J.~P.
  Mellado, T.~Mieslinger, E.~Mohino, H.~Paw{\l}owska, K.~{Peters-von Gehlen},
  A.~Sarr{\'e}, P.~Sobhani, P.~Stier, L.~Tuppi, P.~L. Vidale, I.~Sandu, and
  B.~Stevens.
\newblock {{nextGEMS}}: Entering the era of kilometer-scale {{Earth}} system
  modeling.
\newblock \emph{Geoscientific Model Development}, 18\penalty0 (20):\penalty0
  7735--7761, Oct. 2025.
\newblock ISSN 1991-959X.
\newblock \doi{10.5194/gmd-18-7735-2025}.

\bibitem[Sellar et~al.(2019)Sellar, Jones, Mulcahy, Tang, Yool, Wiltshire,
  O'Connor, Stringer, Hill, Palmieri, Woodward, {de Mora}, Kuhlbrodt, Rumbold,
  Kelley, Ellis, Johnson, Walton, Abraham, Andrews, Andrews, Archibald,
  Berthou, Burke, Blockley, Carslaw, Dalvi, Edwards, Folberth, Gedney,
  Griffiths, Harper, Hendry, Hewitt, Johnson, Jones, Jones, Keeble, Liddicoat,
  Morgenstern, Parker, Predoi, Robertson, Siahaan, Smith, Swaminathan,
  Woodhouse, Zeng, and Zerroukat]{Sellar2019}
A.~A. Sellar, C.~G. Jones, J.~P. Mulcahy, Y.~Tang, A.~Yool, A.~Wiltshire, F.~M.
  O'Connor, M.~Stringer, R.~Hill, J.~Palmieri, S.~Woodward, L.~{de Mora},
  T.~Kuhlbrodt, S.~T. Rumbold, D.~I. Kelley, R.~Ellis, C.~E. Johnson,
  J.~Walton, N.~L. Abraham, M.~B. Andrews, T.~Andrews, A.~T. Archibald,
  S.~Berthou, E.~Burke, E.~Blockley, K.~Carslaw, M.~Dalvi, J.~Edwards, G.~A.
  Folberth, N.~Gedney, P.~T. Griffiths, A.~B. Harper, M.~A. Hendry, A.~J.
  Hewitt, B.~Johnson, A.~Jones, C.~D. Jones, J.~Keeble, S.~Liddicoat,
  O.~Morgenstern, R.~J. Parker, V.~Predoi, E.~Robertson, A.~Siahaan, R.~S.
  Smith, R.~Swaminathan, M.~T. Woodhouse, G.~Zeng, and M.~Zerroukat.
\newblock {{UKESM1}}: {{Description}} and {{Evaluation}} of the {{U}}.{{K}}.
  {{Earth System Model}}.
\newblock \emph{Journal of Advances in Modeling Earth Systems}, 11\penalty0
  (12):\penalty0 4513--4558, 2019.
\newblock ISSN 1942-2466.
\newblock \doi{10.1029/2019MS001739}.

\bibitem[Shan et~al.(2026)Shan, Patel, Sun, Hassanzadeh, Tziperman, and
  Abbot]{shanTestingLimitsSurface2026}
W.~Shan, D.~Patel, Y.~Sun, P.~Hassanzadeh, E.~Tziperman, and D.~S. Abbot.
\newblock Testing the {{Limits}} of {{Surface QG Theory With Deep Learning}}.
\newblock \emph{Journal of Geophysical Research: Machine Learning and
  Computation}, 3\penalty0 (5):\penalty0 e2026JH001410, 2026.
\newblock ISSN 2993-5210.
\newblock \doi{10.1029/2026JH001410}.

\bibitem[Shcherbina et~al.(2013)Shcherbina, D'Asaro, Lee, Klymak, Molemaker,
  and McWilliams]{shcherbinaStatisticsVerticalVorticity2013}
A.~Y. Shcherbina, E.~A. D'Asaro, C.~M. Lee, J.~M. Klymak, M.~J. Molemaker, and
  J.~C. McWilliams.
\newblock Statistics of vertical vorticity, divergence, and strain in a
  developed submesoscale turbulence field.
\newblock \emph{Geophysical Research Letters}, 40\penalty0 (17):\penalty0
  4706--4711, 2013.
\newblock ISSN 1944-8007.
\newblock \doi{10.1002/grl.50919}.

\bibitem[Silva et~al.(2021)Silva, Ma, Hardin, and
  Rothenberg]{silvaPhysicallyRegularizedMachine2021}
S.~J. Silva, P.-L. Ma, J.~C. Hardin, and D.~Rothenberg.
\newblock Physically regularized machine learning emulators of aerosol
  activation.
\newblock \emph{Geoscientific Model Development}, 14\penalty0 (5):\penalty0
  3067--3077, May 2021.
\newblock ISSN 1991-959X.
\newblock \doi{10.5194/gmd-14-3067-2021}.

\bibitem[Storer and Aluie(2023)]{storerFlowSieveCoarseGrainingUtility2023}
B.~A. Storer and H.~Aluie.
\newblock {{FlowSieve}}: {{A Coarse-Graining Utility}} for {{Geophysical
  Flows}} on the {{Sphere}}.
\newblock \emph{Journal of Open Source Software}, 8\penalty0 (84):\penalty0
  4277, Apr. 2023.
\newblock ISSN 2475-9066.
\newblock \doi{10.21105/joss.04277}.

\bibitem[Storkey et~al.(2025)Storkey, Mathiot, Bell, Copsey, Guiavarc'h,
  Hewitt, Ridley, and Roberts]{storkeyResolutionDependenceInterlinked2025}
D.~Storkey, P.~Mathiot, M.~J. Bell, D.~Copsey, C.~Guiavarc'h, H.~T. Hewitt,
  J.~Ridley, and M.~J. Roberts.
\newblock Resolution dependence of interlinked {{Southern Ocean}} biases in
  global coupled {{HadGEM3}} models.
\newblock \emph{Geoscientific Model Development}, 18\penalty0 (9):\penalty0
  2725--2745, May 2025.
\newblock ISSN 1991-959X.
\newblock \doi{10.5194/gmd-18-2725-2025}.

\bibitem[Subich et~al.(2025)Subich, Husain, Separovic, and
  Yang]{subichFixingDoublePenalty2025}
C.~Subich, S.~Z. Husain, L.~Separovic, and J.~Yang.
\newblock Fixing the {{Double Penalty}} in {{Data-Driven Weather Forecasting
  Through}} a {{Modified Spherical Harmonic Loss Function}}, May 2025.

\bibitem[Uchida et~al.(2023)Uchida, Rokem, Soler, Nicholas, Abernathey,
  {nouguier}, Busecke, Vanderplas, Martin, Mondal, Mayer, Halchenko, Wilson,
  Constantinou, Ponte, Squire, Spring, Pak, RichardScottOZ, Hoyer, and
  Moon]{uchidaXgcmXrftV1012023}
T.~Uchida, A.~Rokem, S.~Soler, T.~Nicholas, R.~Abernathey, f.~{nouguier},
  J.~Busecke, J.~Vanderplas, P.~Martin, A.~Mondal, A.~Mayer, Y.~Halchenko,
  G.~Wilson, N.~C. Constantinou, A.~Ponte, D.~Squire, A.~Spring, K.~Pak,
  RichardScottOZ, S.~Hoyer, and Z.~Moon.
\newblock Xgcm/xrft: V1.0.1.
\newblock Feb. 2023.
\newblock \doi{10.5281/zenodo.7621857}.

\bibitem[Wenegrat et~al.(2026)Wenegrat, Chor, and
  Barkan]{wenegratCoarsegrainedLocalAvailable2026}
J.~O. Wenegrat, T.~Chor, and R.~Barkan.
\newblock Coarse-grained local available potential energy, May 2026.

\bibitem[Wilder(2026{\natexlab{a}})]{wilderFiguresFigureData2026}
T.~Wilder.
\newblock Figures and figure data for the paper: {{Evaluating}} the prediction
  of subgrid eddy kinetic energy from a machine learning model.
\newblock Oct. 2026{\natexlab{a}}.
\newblock \doi{10.5281/zenodo.23814762}.

\bibitem[Wilder(2026{\natexlab{b}})]{wilderMlparameterisation2026}
T.~Wilder.
\newblock Ml-parameterisation.
\newblock Oct. 2026{\natexlab{b}}.
\newblock \doi{10.5281/zenodo.23815035}.

\bibitem[Wilder(2026{\natexlab{c}})]{wilderTrainingDataModels2026}
T.~Wilder.
\newblock Training data, models and predictions for the paper: {{Evaluating}}
  the prediction of subgrid eddy kinetic energy from a machine learning model.
\newblock Oct. 2026{\natexlab{c}}.
\newblock \doi{10.5281/zenodo.23814760}.

\bibitem[Xie et~al.(2025)Xie, Gao, and Lu]{xieGlobalEstimationSubsurface2025}
C.~Xie, A.-K. Gao, and X.~Lu.
\newblock Global {{Estimation}} of {{Subsurface Eddy Kinetic Energy}} of
  {{Mesoscale Eddies Using Machine Learning}}.
\newblock \emph{Journal of Geophysical Research: Machine Learning and
  Computation}, 2\penalty0 (3):\penalty0 e2024JH000578, 2025.
\newblock ISSN 2993-5210.
\newblock \doi{10.1029/2024JH000578}.

\bibitem[Yan et~al.(2024)Yan, Mak, and Wang]{yanChoiceTrainingData2024}
F.~E. Yan, J.~Mak, and Y.~Wang.
\newblock On the {{Choice}} of {{Training Data}} for {{Machine Learning}} of
  {{Geostrophic Mesoscale Turbulence}}.
\newblock \emph{Journal of Advances in Modeling Earth Systems}, 16\penalty0
  (2):\penalty0 e2023MS003915, 2024.
\newblock ISSN 1942-2466.
\newblock \doi{10.1029/2023MS003915}.

\bibitem[Zanna and Bolton(2020)]{zannaDataDrivenEquationDiscovery2020}
L.~Zanna and T.~Bolton.
\newblock Data-{{Driven Equation Discovery}} of {{Ocean Mesoscale Closures}}.
\newblock \emph{Geophysical Research Letters}, 47\penalty0 (17):\penalty0
  e2020GL088376, 2020.
\newblock ISSN 1944-8007.
\newblock \doi{10.1029/2020GL088376}.

\bibitem[Zanna and Bolton(2021)]{ZannaBolton2021}
L.~Zanna and T.~Bolton.
\newblock Deep {{Learning}} of {{Unresolved Turbulent Ocean Processes}} in
  {{Climate Models}}.
\newblock In \emph{Deep {{Learning}} for the {{Earth Sciences}}}, chapter~20,
  pages 298--306. John Wiley \& Sons, Ltd, 2021.
\newblock ISBN 978-1-119-64618-1.
\newblock \doi{10.1002/9781119646181.ch20}.

\bibitem[Zanna et~al.(2025)Zanna, Gregory, Perezhogin, Sane, Zhang, Adcroft,
  Bushuk, {Fernandez-Granda}, Reichl, Balwada, Busecke, Chapman, Connolly, Du,
  Everard, Falasca, Falga, Kamm, Meunier, Liu, Nasser, Pudig, Shao, Simpson,
  Vogt, and Wu]{zannaFrameworkHybridPhysicsAI2025}
L.~Zanna, W.~Gregory, P.~Perezhogin, A.~Sane, C.~Zhang, A.~Adcroft, M.~Bushuk,
  C.~{Fernandez-Granda}, B.~Reichl, D.~Balwada, J.~Busecke, W.~Chapman,
  A.~Connolly, D.~Du, K.~Everard, F.~Falasca, R.~Falga, D.~Kamm, E.~Meunier,
  Q.~Liu, A.~Nasser, M.~Pudig, A.~Shao, J.~L. Simpson, L.~Vogt, and J.~Wu.
\newblock A {{Framework}} for {{Hybrid Physics-AI Coupled Ocean Models}}, Oct.
  2025.

\bibitem[Zhang et~al.(2025)Zhang, Perezhogin, Adcroft, and
  Zanna]{zhangAddressingOutofSampleIssues2025}
C.~Zhang, P.~Perezhogin, A.~Adcroft, and L.~Zanna.
\newblock Addressing {{Out}}-of-{{Sample Issues}} in {{Multi-Layer
  Convolutional Neural-Network Parameterization}} of {{Mesoscale Eddies Applied
  Near Coastlines}}.
\newblock \emph{Journal of Advances in Modeling Earth Systems}, 17\penalty0
  (5):\penalty0 e2024MS004819, 2025.
\newblock ISSN 1942-2466.
\newblock \doi{10.1029/2024MS004819}.

\bibitem[Zhuang et~al.(2025)Zhuang, {dussin}, Bourgault, Huard, Banihirwe,
  Raynaud, Malevich, Schupfner, Gauthier, Filipe, Levang, Almansi, J{\"u}ling,
  RichardScottOZ, RondeauG, Rasp, Smith, Meyer, Mares, Sacks, Stachelek,
  Plough, Pierre, Bell, Caneill, and Li]{zhuangPangeodataXESMFV0922025}
J.~Zhuang, r.~{dussin}, P.~Bourgault, D.~Huard, A.~Banihirwe, S.~Raynaud,
  B.~Malevich, M.~Schupfner, C.~Gauthier, Filipe, S.~Levang, M.~Almansi,
  A.~J{\"u}ling, RichardScottOZ, RondeauG, S.~Rasp, T.~J. Smith, A.~G. Meyer,
  B.~Mares, B.~Sacks, J.~Stachelek, M.~Plough, Pierre, R.~Bell, R.~Caneill, and
  X.~Li.
\newblock Pangeo-data/{{xESMF}}: V0.9.2 - {{Third}} time's a charm.
\newblock Nov. 2025.
\newblock \doi{10.5281/zenodo.17743552}.

\end{thebibliography}
